\documentclass[reprint,amsmath,twocolumn,amssymb, nobibnotes, aps, pre, superscriptaddress]{revtex4-1}

\usepackage{graphicx}
\usepackage{dcolumn}
\usepackage{bm}
\usepackage{color}
\usepackage{tikz}
\usetikzlibrary{shapes}
\usetikzlibrary{backgrounds}
\usetikzlibrary{plotmarks}

\begin{document}
\def \beq{\begin{equation}}
\def \eeq{\end{equation}}
\def \bea{\begin{eqnarray}}
\def \eea{\end{eqnarray}}
\def \bem{\begin{displaymath}}
\def \eem{\end{displaymath}}
\def \P{\Psi}
\def \Pd{|\Psi(\boldsymbol{r})|}
\def \Pds{|\Psi^{\ast}(\boldsymbol{r})|}
\def \Po{\overline{\Psi}}
\def \bs{\boldsymbol}
\def \bl{\bar{\boldsymbol{l}}}
\newcommand{\ihat}{\hat{\textbf{\i}}}
\newcommand{\jhat}{\hat{\textbf{\j}}}

\title{ {Topological insulators in longitudinally driven waveguides: \\Lieb and Kagome lattices}}
\author{Mark J. Ablowitz and Justin T. Cole}
\affiliation{Department of Applied Mathematics, University of Colorado, Boulder, Colorado 80309}
\date{\today}     
\begin{abstract}

{Topological insulators are studied via tight-binding approximations of} 
longitudinally driven photonic lattices with three lattice sites per unit cell. Two cases {are} considered in detail: Lieb and Kagome lattices. The lattice is decomposed into three sublattices each of which {are allowed} move independently of one another. Emphasis is placed on periodic driving induced by laser-etched 
helical coils along the direction of propagation. The linear Floquet bands are constructed for various {inter-}sublattice rotation patterns such as:  different radii, different frequency, {phase offset and} quasi one-dimensional motion. Depending on the nature of the band structure, {bulk spectral bands with nonzero Chern number are found to support topologically protected} edge states {which can} move unidirectionally. In this case, {the modes} move {scatter-free around} 
defects due to underlying topological protection. { {Intriguing mode dynamics are found including bi-directional topological modes and} bulk-edge leakage i.e. excitation of bulk modes at a defect for edge modes with dispersion frequencies nearby the bulk bands. Finally, certain nonlinear edge modes are {also}  found to propagate unidirectionally and scatter-free around lattice defects. }



\begin{description}

\item[PACS numbers] 42.65.Tg, 42.65.Jx, 42.82.Et
\end{description}
\end{abstract}
\pacs{Valid PACS appear here}
\maketitle

\section{Introduction}
\label{intro}

{The study of photonic topological insulators has received} considerable interest due in part to their remarkable mode propagation properties {and special connection to} {underlying topology}. 
In photonic systems, breaking time-reversal symmetry has been shown to support topologically protected states \cite{HaldRag2008}. {The bulk bands associated with these systems possess nonzero Chern invariants  and, as a result, {exhibit} {so-called} gapless bands}. {Physically,} the corresponding modes manifest themselves in edge states that {can} travel in one direction and {propagate stably and without scatter {at} {boundary} 
defects.}

The first {experimental} realization of a topologically protected electromagnetic wave was observed in \cite{Wang2009}. In that work an external magnetic field was applied to a periodic array of ferrite rods in order to break time-reversal symmetry. {There it was shown that microwaves could be localized along the device boundary, move unidirectionally, and  propagate} scatter-free around barrier defects.

 {In {the {photonic} regime,} 
 a {magnet-free} experimental} realization of a topological insulator was  presented in \cite{RechtsSegev2013}. In {that} work time-reversal symmetry was broken by a honeycomb array of helically-varying waveguides etched {into} a bulk medium. In spatial optics the direction of propagation plays the role of time and  so the helical variations act as a time-dependent potential. 
This paper {also} focuses on a system {of helically-varying waveguides, {however} here we {consider more complicated geometries.}



Longitudinally} varying {waveguide} arrays {have} been used to explore numerous other topological {effects.} 
Linear and nonlinear staggered square arrangements were considered in \cite{Leykam2016A} and \cite{Leykam2016}. By introducing a phase offset among the sublattice waveguides it is possible to observe a {topological} 
transition point {known} 
as a conical Weyl point \cite{Noh2017}. For intense nonlinear beams the edge mode envelope has been found to be governed by a nonlinear Schr\"odinger equation and, in certain parameter regimes, support topologically protected edge solitons \cite{AbCo2017} as well as exhibit modulational instability \cite{Lumer2016}. Unidirectional edge modes have also been observed to propagate in helically-driven quasicrystals \cite{Bandres2016}; {i.e.} structures that are {not periodic} but contain some long range order.

{Applications for these type of systems} {have been explored.}  Photonic topological systems {have {the} potential to act as}
optical isolators (one-way transmitters) \cite{Zhou2017, ElGanainy2015} and circulators \cite{Qiu2011}{. ~
 {Different types of photonic topological insulator systems} have also been found to generate and support {novel} {and} robust laser systems \cite{Bahari2017,StJean2017,Parto2017,Bandres2018,Harari2018}.
 {Properties and evolution of {high power} nonlinear 
 states were investigated in \cite{AbCo2017} and \cite{AbCuMa2014}.}

{In \cite{AbCo2017} a systematic approach to derive tight-binding models in honeycomb and {staggered} square lattices (two lattice sites per unit cell) was developed.} This paper extends those methods to more complex (three lattice sites per unit cell) periodically driven arrays. In particular, we focus on the Lieb and Kagome lattices. {These are two of the simplest {2d} lattices possible and they have come up in the context of many important physical systems.
Lieb lattices have been used in optical settings (see below) as well as {in the} study topological spin states of cold fermionic systems \cite{Goldman2011}. The Kagome lattice has been used as a tunable optical lattice for ultracold atoms \cite{Jo2012}, {and also occurs} in materials that are candidates for observing quantum spin liquids \cite{Helton2010}.}

In the absence of external driving {both the Lieb and Kagome} lattices can support localized flat band modes in the bulk \cite{GuzmanSilva2014,Vicencio2015,Mukherjee2015,Zong2016}, or along the edge \cite{GuzmanSilva2014, Bandres2014}. Interestingly, flat band modes can be diffraction-free \cite{Vicencio2014}. Introducing {uniform} helical variation to a Lieb waveguide array has been shown to offer a rich set of edge mode dynamics \cite{Bandres2014}. These include the presence of a (stationary) flat band mode, as well as (traveling) topologically protected modes. 
Here we generalize the {latter} work 
to allow more complex sublattice rotation patterns. {To our knowledge, edge mode dynamics for helically driven Kagome lattices {have} been {studied} much less than {the} Lieb lattice.}
 {We find driven Kagome lattices lead to novel phenomena such as having traveling topological modes which are not always unidirectional.}


In this work we decompose the Lieb and Kagome lattices into three sublattices that can move independently of {one another} {and}  {frequently} find gapless modes. 
{We derive a tight-binding system, valid in a deep lattice limit, that describes incoming light beams. This model takes into account the lattice driving via periodic functions that are parametrized in terms of the direction of propagation, i.e. the ``time'' variable.} As a result, we {obtain} a system {of differential equations} with periodic coefficients of which we compute the Floquet bands, {their} {corresponding} {eigenmodes/edge modes, and associated topological invariants (Chern numbers)}. {We also briefly consider} {nonlinear edge modes that exhibit properties of {topological} protection.}




\subsection{Paraxial Wave Equation}
\label{paraxial}

The propagation of {intense, paraxial}  light beams in Kerr waveguide arrays is 
governed by the nonlinear Schr\"odinger (NLS) equation
\begin{equation}
\label{NLS}
i \frac{ \partial \psi}{\partial z} + \frac{1}{2 k_0} \nabla^2 \psi - \frac{k_0}{n_0}  \left( n_{\ell}({\bf r},z) - n_2 |\psi|^2 \right)\psi = 0  \; ,
\end{equation}
where  $k_0$ is the carrier wavenumber, $n_0$ is the bulk media index of refraction, and $\nabla^2 \equiv \partial_x^2 + \partial_y^2$. The complex electric field envelope $\psi({\bf r},z)$ depends on the position in both the direction of propagation, $z$, and the transverse plane, ${\bf r} = (x,y)$. Regions of high refractive index are carved into bulk media using a femtosecond laser etching process \cite{SzNo10}. The presence of these lattice waveguides is represented by the potential function $- n_{\ell}({\bf r},z) .$ {Additionally,} focusing Kerr nonlinear media ($n_2 > 0$) exhibits an intensity-dependent response to the incoming light beam. {It is {directly} from equation (\ref{NLS}) that we derive our tight-binding {system}.}

In this paper we restrict our attention to non-simple lattices with three lattice sites ($a,b,c$) per unit cell. To model such a scenario, we rewrite the lattice potential as the combination of three interpenetrating sublattices
\begin{equation}
\label{general_potential}
n_{\ell} ({\bf r},z) = |\Delta n| \bigg[ 1 -  V_a({\bf r}, z) -  V_b({\bf r}, z)  - V_c({\bf r}, z)  \bigg] \; ,
\end{equation}
where $|\Delta n|$ denotes the contrast in waveguide refractive index from {that of} the bulk.
Each sublattice {is taken to consist} of a sum {of} Gaussians 
\begin{align}
\label{sublattice_define}
 V_{j} ({\bf r} ,z) &= \sum_{{\bf v} \in \mathcal{R}_j} \tilde{V}({\bf r} - {\bf v} - {\bf h}_j(z)) \; , ~ j = a,b,c
\\ \nonumber
 & \tilde{V}({\bf r}) =  \exp{\left( - \frac{ x^2 }{\sigma_x^2} - \frac{y^2}{\sigma_y^2} \right)} \; ,
\end{align}
{where} $ \sigma_x, \sigma_y > 0 $. {The set $\mathcal{R}_j$ consists of all the lattice sites for the $j^{\rm th}$ sublattice.} 


 The minima of the lattice potential correspond to the center of the waveguides where the index of refraction is largest. The smooth parametric functions ${\bf h}_j(z)$ drive the sublattices. The parameters $\sigma_x,\sigma_y$ control the geometric shape of the waveguides. When $\sigma_x = \sigma_y$ the waveguides are circular (isotropic), whereas the waveguides are elliptical (anisotropic) when $\sigma_x \not= \sigma_y$.

{Here we} concentrate on driving functions which are periodic in $z$.
Specifically, we examine 
\begin{equation}
{\bf h}_{j}(z) = \tilde{R}_j \left( \cos\left( \Lambda_j z + \chi_j \right) , \sin\left( \Lambda_j z + \chi_j \right) \right) \; , 
\end{equation}
where $\tilde{R}_j$ is the helix radius, {$\Lambda_j$} is the angular frequency, and $\chi_j$ is an arbitrary phase shift.
 {It is useful to} {transform to a}  coordinate frame co-moving with the $V_b({\bf r}, z)$ sublattice by performing the change of variable
\begin{equation*}
\tilde{{\bf r}} = {\bf r}  - {\bf h}_b(z) ~ , ~~~ \tilde{z} = z \; .
\end{equation*}
 {Introducing the transformation}
\begin{equation*}
\psi({\bf r},z) ={\tilde{\psi}}(\tilde{{\bf r}},\tilde{z}) \exp\left(  \frac{ i  \int_0^{\tilde{z}} | {\bf A}(\zeta)|^2 d \zeta }{2 k_0} \right) \; ,
\end{equation*}
for the pseudo-field (vector potential) 
\begin{equation*}
 {\bf A}(\tilde{z}) = - k_0 {\bf h}'_b(\tilde{z}) \; ,
\end{equation*}
{yields} (after dropping the tilde notation)
\begin{equation}
\label{LSE_3}
i \frac{ \partial \psi}{\partial z} + \frac{1}{2 k_0} \left( \nabla + i  {\bf A}(z)\right)^2 \psi - \frac{k_0}{n_0}  \left( n_{\ell}({\bf r},z) -  n_2 |\psi|^2 \right)\psi = 0  \; .
\end{equation}
This equation is nondimensionalized by
\begin{align*}
& x = \ell x' ~ , ~ y = \ell y' ~ , ~ z = z_* z' ~ , \\
 \sigma_x & =  \ell \sigma_x' ~, ~ \sigma_y = \ell \sigma_y' ~ , ~ \psi = \sqrt{I_*} \psi' ~ , 
\end{align*}
where $\ell$ is the distance between nearest neighbor lattice sites, $z_* = 2 k_0 \ell^2$ is {a} typical propagation distance, {and} $I_*$ is the peak intensity of the input beam. Dropping the {prime} notation {gives} the dimensionless equation
\begin{equation}
\label{LSE_4}
i \frac{ \partial \psi}{\partial z} + \left( \nabla + i {\bf A}(z)\right)^2 \psi - V({\bf r},z) \psi + \gamma |\psi|^2 \psi = 0  \; ,
\end{equation}
{where $\gamma=2 k_0^2 \ell^2 n_2I_*/n_0 \ge 0$,}
with potential function
\begin{equation*}
\label{potl}
V({\bf r},z)  = V_0^2 \bigg[ 1 -  V_a({\bf r} - \Delta {\bf h}_{ab}(z)) -  V_b({\bf r} )-  V_c({\bf r} - \Delta {\bf h}_{cb}(z)) \bigg] \; ,
\end{equation*}
{which has amplitude $V_0^2 =2k_0^2\ell^2 |\Delta n| /n_0$, and}
\begin{equation}
\Delta {\bf h}_{ij}(z) \equiv {\bf h}_i(z) - {\bf h}_j(z) ~~~~~ i,j= a,b,c \; . 
\end{equation}
In dimensionless form, the driving functions are 
\begin{equation*}
 {\bf h}_j(z) = \eta_j \left( \cos\left( \Omega_j z + \chi_j \right) , \sin\left( \Omega_j z + \chi_j \right) \right) \; ,  ~ j = a,b,c
\end{equation*}
where $\eta_j = \tilde{R}_j / \ell$ and $ \Omega_j = \Lambda_j z_*.$



Finally, we introduce the phase transformation
\begin{equation*}
 \psi({\bf r},z) = \phi({\bf r},z) e^{ - i {\bf r} \cdot {\bf A}(z)}  \; ,
\end{equation*}
which {simplifies} {Eq.~(\ref{LSE_4}) to}
\begin{equation}
\label{LSE_5}
i \frac{ \partial \phi}{\partial z} +  \nabla^2 \phi + {\bf r} \cdot {\bf A}_{z} \phi - V({\bf r},z) \phi + \gamma |\phi|^2 \phi = 0  \; .
\end{equation}
The pseudo-field in dimensionless coordinates is given by
\begin{equation}
\label{define_pseudo_field}
{\bf A}(z) = \kappa \left( \sin \left( \Omega_b z  + \chi_b \right) , - \cos \left( \Omega_b z  + \chi_b \right) \right) \; ,
\end{equation}
(since we are working in the ${\bf h}_b(z)$ reference frame) where {$\kappa = k_0 \ell \tilde{R}_b \Lambda_b $.} 

\section{Tight-binding Approximation}
\label{TBA_derive}

We now {develop} a tight-binding approximation {from} Eq.~(\ref{LSE_5}). This {discrete} 
model {assumes 
a sharp} contrast {of index of refraction between the waveguides and bulk background. 
Put another way,} 
we are working in a deep lattice regime{; i.e. $V_0^2 \gg 1$. In addition, locally} we approximate the lattice potential by a 
paraboloid potential {in order to obtain an orbital approximation.
The beam field $\phi({\bf r}, z)$ is expressed as a sum of these {strongly decaying} 
orbital functions centered at the ($z$-dependent) lattice sites.} The localized orbital modes are taken to satisfy a harmonic oscillator equation with {a} quadratic potential 
that is periodic in $z$. 
In the {deep lattice limit} 
these 
 {orbital} {modes} are well-localized so that the only significant interactions occur between nearest neighbors. 

Near the well-separated 
 {potential} minima we approximate the {sublattices given in (\ref{sublattice_define})} 
by the first few terms of their Taylor series. The local approximation of the potential is the paraboloid 
\begin{equation}
\label{paraboloid_pot}
 \overline{V} ({\bf r}) = V_0^2 \bigg( \frac{x^2}{\sigma_x^2} +    \frac{y^2}{\sigma_y^2} \bigg)  \; .
\end{equation}
Near {here} 
the wave field is taken to satisfy the orbital equation
\begin{equation}
\left[ - \nabla^2   + \overline{V}({\bf r} - {\bf v}^j - \Delta {\bf h}_{jb}(z)) \right] \phi_{j,{\bf v}} = E  \phi_{j,{\bf v}}\; ,  ~ j = a,b,c 
\end{equation}
 for eigenvalue {$E = V_0(1/\sigma_x + 1/\sigma_y)$} and {the normalized} {Gaussian} {eigenfunction 
 \begin{align}
 \label{orbital_formula}
 &\phi_{j,{\bf v}}({\bf r},z) = \sqrt{\frac{V_0}{\pi \sqrt{\sigma_x \sigma_y}}} \times \\ \nonumber
 & \exp\left[ - \frac{V_0 [{\bf r} - {\bf v}^j  - \Delta {\bf h}_{jb}(z) ]^2_x}{ 2 \sigma_x} - \frac{V_0 [{\bf r} - {\bf v}^j - \Delta {\bf h}_{jb}(z) ]^2_y}{ 2 \sigma_y} \right]   \; ,
 \end{align}
 where the subscripts $x$ and $y$ denote the first and second components, respectively {and ${\bf v}$ denotes the lattice location.}
 The eigenvalue $E$ is the same for each sublattice as long as $V_0, \sigma_x$ and $\sigma_y$ are identical in {all}  sublattices.
 

{The wave field is decomposed into a sum} 
of {the above} orbital functions, given by {
\begin{align}
\label{ansatz_define}
\phi({\bf r},z) =  \sum_{m,n} &\big[ a_{mn}(z) \phi_{a,{\bf v}_{mn}}({\bf r} , z) + b_{mn}(z) \phi_{b,{\bf v}_{mn}}({\bf r}) \\ \nonumber
+&~ c_{mn}(z) \phi_{c,{\bf v}_{mn}}({\bf r},z)  \big] e^{- i E z} \; ,
\end{align}
}where the orbitals are modulated by the associated coefficients {$a_{mn},b_{mn},c_{mn}.$}
We substitute expansion (\ref{ansatz_define}) into (\ref{LSE_5}), multiply by {$\phi_{i,{\bf p}}({\bf r},z), i = a,b,c$}, and integrate {over all ${\bf r}$}. {We {label} the lattice the manner shown in Figs.~\ref{lieb_fig} or \ref{kagome_fig}. 
The details of a similar derivation are described in \cite{AbCo2017}.

The ansatz in (\ref{ansatz_define}) is similar to that used in Wannier functions  \cite{Marzari2012}. Wannier modes are the Fourier coefficients of a linear Bloch wave and they exponentially localized at the lattice sites. 
One drawback to using actual Wannier functions is that ({usually}) there is no {convenient analytical formulae} and they must be computed numerically. Approximating the Bloch functions by the orbitals allows us to derive coefficients that depend directly on the physical parameters of waveguide system (\ref{LSE_5}). {Furthermore, they} capture the appropriate band dynamics below.}



The two lattices we consider in detail, Lieb and Kagome, are shown in Figs.~\ref{lieb_fig} and \ref{kagome_fig}, respectively. Both are nondimensionalized so that the distance between nearest neighbors is one.
The {lattice sites of the} Lieb lattice 
are separated via the standard basis vectors
\begin{equation}
\label{lieb_lattice_vec_define}
{\bf e}_1 = \left( 1 ,  0 \right) , ~~~~  {\bf e}_2 = \left( 0, 1\right) \; .
\end{equation}
The {nearest neighbor distances in the} Kagome lattice {are} defined in terms of the vectors
\begin{equation}
\label{kagome_lattice_vec_define}
{\bf v}_1 = \left(  \frac{\sqrt{3}}{2}, \frac{1}{2} \right) , ~~  {\bf v}_2 =\left( \frac{\sqrt{3}}{2} ,- \frac{1}{2}\right) \; , ~~  {\bf v}_3 =\left( 0,1 \right) \; .
\end{equation}
 {{As a result, the Lieb lattice sites used are given by the sets $\mathcal{R}_a = \{ {\bf v}^a_{mn} \mid {\bf v}^a_{mn} = 2m {\bf e}_2 + (2 n - 1) {\bf e}_1 \}$, $\mathcal{R}_b = \{ {\bf v}^b_{mn} \mid {\bf v}^b_{mn} = 2m {\bf e}_2 + 2 n{\bf e}_1 \}$, $\mathcal{R}_c = \{ {\bf v}^c_{mn} \mid {\bf v}^c_{mn} = ( 2 m +1 ){\bf e}_2 + 2n{\bf e}_1 \}$ for $m,n \in \mathbb{Z}$.} The Lieb (periodic) lattice vectors are $2{\bf e}_1$ and $2 {\bf e}_2$.

{In}  the case of the Kagome lattice the lattice points are {located at} $\mathcal{R}_a = \{ {\bf v}^a_{mn} \mid {\bf v}^a_{mn} =  (2 m+1) {\bf v}_1 +  2 n {\bf v}_2 \}$, $\mathcal{R}_b = \{ {\bf v}^b_{mn} \mid {\bf v}^b_{mn} = 2m{\bf v}_1 + 2n {\bf v}_2 \}$, $\mathcal{R}_c = \{ {\bf v}^c_{mn} \mid {\bf v}^c_{mn} = 2m {\bf v}_1 + (2n +1) {\bf v}_2 \}$ where  $m,n \in \mathbb{Z}$.  The Kagome (periodic) lattice vectors are $2{\bf v}_1$ and $2 {\bf v}_2$. To index the Kagome lattice we rewrite the labels in terms of the vectors ${\bf w}_1 = (0,1)$ and ${\bf w}_2 = (\sqrt{3},0)$ which are related to the lattice vectors by the linear transformation: $2 {\bf v}_1 = {\bf w}_1 + {\bf w}_2$ and $2 {\bf v}_2 = {\bf w}_2 - {\bf w}_1$; hence {we introduce} the new labels $\tilde{m} = m - n$ and $\tilde{n} = m + n$ (the tilde notation is dropped after this point{; so the equations for the Kagome lattice below are in the ${\bf w}_1, {\bf w}_2$ directions}). Doing this makes it easier to find edge modes along the left or right edges. {All points in the Lieb/Kagome lattices are obtained from translations of the lattice vectors beginning with three initial points.}

\begin{figure}
\centering
\includegraphics[scale=.69]{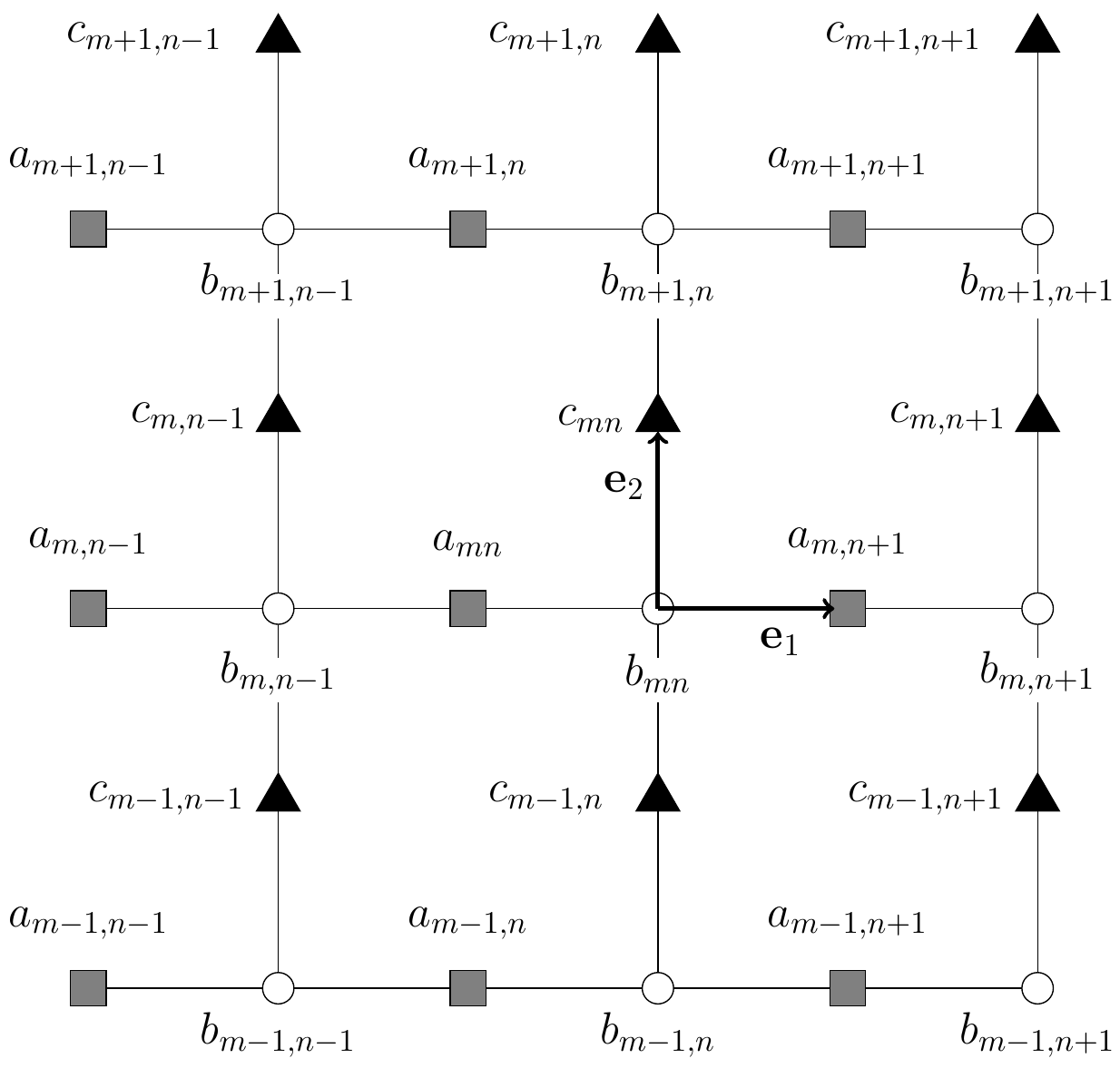}
 \caption{The Lieb lattice consists of three interpenetrating square sublattices $V_a({\bf r})$ (square site, ${\bf a}$), $V_b({\bf r})$ (circle site, ${\bf b}$) and $V_c({\bf r})$ (triangle site, ${\bf c}$). The vectors ${\bf e}_1$ and ${\bf e}_2$ are defined in (\ref{lieb_lattice_vec_define}). Lines denote nearest neighbor interactions. Shown is a bearded (straight) boundary condition on the left (right) edge.  \label{lieb_fig}}
\end{figure}

To leading order: in the Lieb lattice the $b$ sites interact with both the nearest $a$ and $c$ sites, whereas the $a,c$ sites 
only interact with the nearest $b$ sites; in the Kagome lattice all nearest sites interact with {the others}.
For experimental parameters used in \cite{RechtsSegev2013} the coefficients {in Eqs.~(\ref{lieb_TBA_eq1})-(\ref{lieb_TBA_eq3}) and (\ref{kagome_TBA_eq1})-(\ref{kagome_TBA_eq3})} that represent nearest neighbor interaction 
are on the order of $\mathcal{O}(10^{-1})$, while the next-nearest neighbor coupling {(not included)} is on the order of $\mathcal{O}({10}^{-3})$ for Lieb and $\mathcal{O}({10}^{-5})$ for Kagome.
For this reason, we only consider nearest neighbor overlap, but emphasize that longer range interactions could also be taken into account.


\begin{figure}
\centering
\includegraphics[scale=.52]{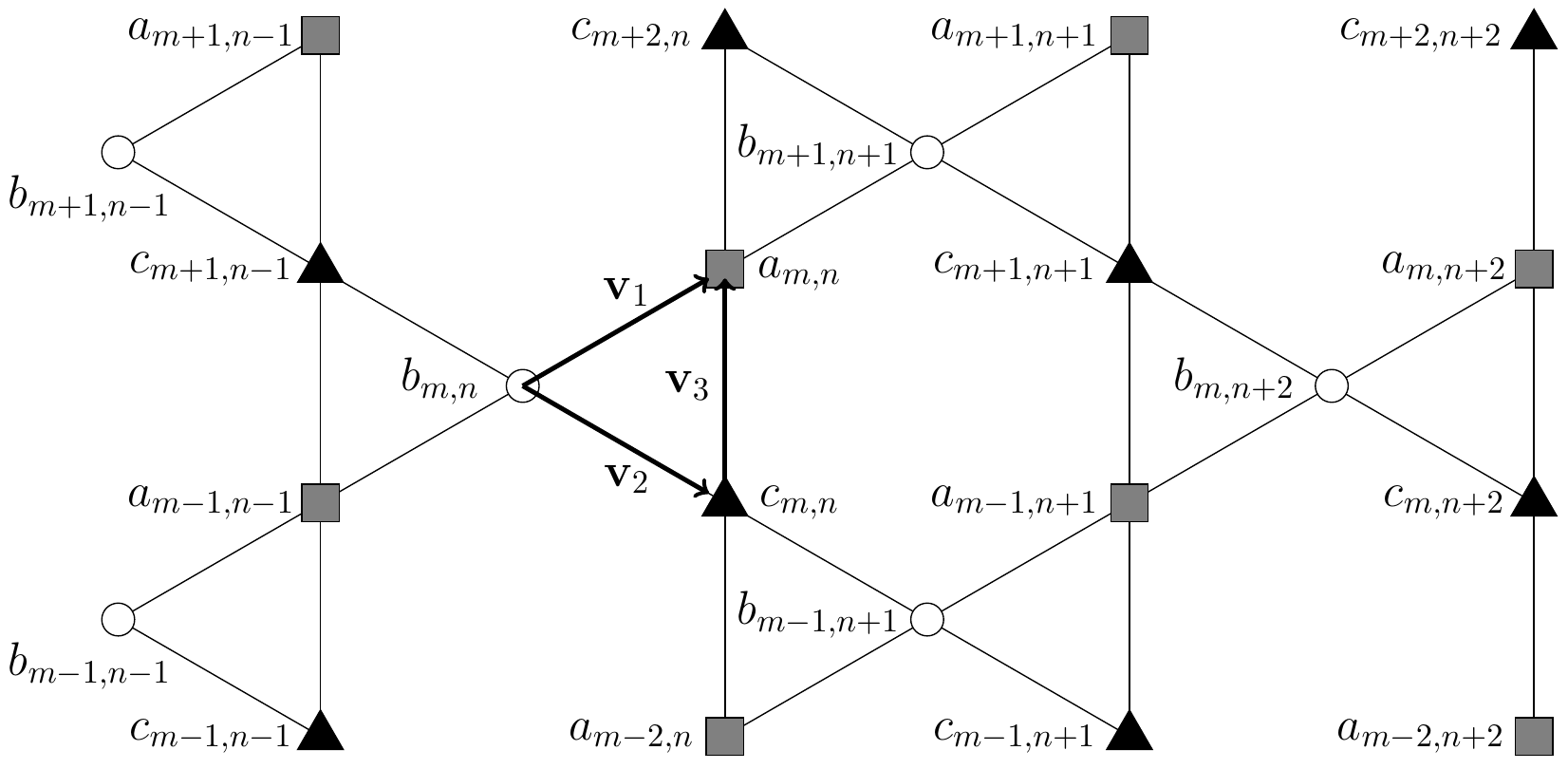}
 \caption{The Kagome lattice consists of three interpenetrating triangular sublattices $V_a({\bf r})$ (square sites, ${\bf a}$), $V_b({\bf r})$ (circle sites, ${\bf b}$) and $V_c({\bf r})$ (triangle sites, ${\bf c}$). The  vectors ${\bf v}_1, {\bf v}_2$, and ${\bf v}_3$ are given in (\ref{kagome_lattice_vec_define}). Lines denote nearest neighbor interactions.   Shown is a pointy (straight) boundary condition on the left (right) edge. \label{kagome_fig}}
\end{figure}


\subsection{Lieb Tight-binding Approximation}
\label{lieb_TBA}

Here we present a tight-binding model for the Lieb lattice. 
In deriving this set of equations only the dominant self and nearest neighbor interactions are taken into account. 
Since this is a leading order calculation we do not include any direct interaction between the $a$ and $c$ lattice sites (see Fig.~\ref{lieb_fig}) in our derivation. Weak on-site cubic nonlinearity is included in the equations. The semi-discrete system of {coupled mode} equations are {given by}  





{
\begin{align}
 \label{lieb_TBA_eq1}
 i \frac{d a_{mn}}{dz} &+ ( \Delta {\bf h}_{ab} \cdot {\bf A}_z + \gamma_{\rm nl} |a_{mn}|^2 ) a_{mn} \\ \nonumber
&+ \mathbb{L}_1^{ab}(z) b_{mn} + \mathbb{L}_{-1}^{ab}(z) b_{m,n-1} = 0 \; ,
\end{align}
\begin{align}
 \label{lieb_TBA_eq2}
 i \frac{d b_{mn}}{dz} &+ \gamma_{\rm nl} |b_{mn}|^2 b_{mn} \\ \nonumber
 & +  \mathbb{L}_1^{ba}(z) a_{m,n+1} + \mathbb{L}_{-1}^{ba}(z) a_{mn} \\ \nonumber
  & + \mathbb{L}_2^{bc}(z) c_{mn} + \mathbb{L}_{-2}^{bc}(z) c_{m-1,n} = 0 \; ,
\end{align}
\begin{align}
\label{lieb_TBA_eq3}
  i  \frac{d c_{mn}}{dz} & +  ( \Delta {\bf h}_{cb} \cdot {\bf A}_z + \gamma_{\rm nl} |c_{mn}|^2 )c_{mn} \\  \nonumber
& + \mathbb{L}_2^{cb}(z) b_{m+1,n} + \mathbb{L}_{-2}^{cb}(z) b_{mn} = 0 \; ,
\end{align}
such that $\mathbb{L}_{\pm \ell}^{ij}(z) = \mathbb{L}(\pm {\bf e}_\ell - \Delta {\bf h}_{ij}(z))$} and  {$\gamma_{\rm nl} \ge 0$}.
The definitions of the coefficients  $\mathbb{L}$ and {$\gamma_{\rm nl}$}, in terms of physical parameters, are given in Appendix \ref{TBA_coefficients}. Note that Eq.~(\ref{lieb_TBA_eq2}), unlike the other two equations, does not contain {a linear coefficient depending on ${\bf A}_z$ since $\Delta {\bf h}_{bb}  = 0.$} 


We concentrate on edge modes that propagate along a {semi-infinite strip.} Outside this 
region the light beam is assumed to be negligibly small, {hence} we take zero boundary conditions on the left and right sides. Along the infinite $m$-direction we take large computational domains and implement periodic boundary conditions. For the lattice displayed in Fig.~\ref{lieb_fig} we show two possible edge types: ``bearded" (left) and ``straight" (right). Any combination of these two edge types is possible e.g. straight-straight, straight-bearded, or bearded-bearded.

To find edge mode solutions localized along the left or right boundaries we look for solutions of the form {
\begin{align}
\nonumber
  a_{mn}(z) &= a_{n}( k_y; z) e^{i  2m k_y   }  \; , \\ 
  b_{mn}(z) &= b_{n}( k_y; z) e^{i 2m k_y }  \; , \\ \nonumber
  c_{mn}(z) &= c_{n}( k_y; z) e^{i2m k_y } \; , 
\end{align}
}which reduce Eqs.~(\ref{lieb_TBA_eq1})-(\ref{lieb_TBA_eq3}) to
{
\begin{align}
\label{1d_lieb_TBA_eq1}
 i \frac{d a_{n}}{dz} + & (\Delta {\bf h}_{ab} \cdot {\bf A}_z +  \gamma_{\rm nl} |a_{n}|^2 )a_{n} \\ \nonumber
+ & \mathbb{L}_1^{ab}(z) b_{n} + \mathbb{L}_{-1}^{ab}(z) b_{n-1} = 0 \; ,  
\end{align}
\begin{align}
 \label{1d_lieb_TBA_eq2}
 i \frac{d b_{n}}{dz} &+ \gamma_{\rm nl} |b_{n}|^2 b_{n} \\ \nonumber
 & +  \mathbb{L}_1^{ba}(z) a_{n+1} + \mathbb{L}_{-1}^{ba}(z) a_{n} \\ \nonumber
  & + \mathbb{L}_2^{bc}(z) c_{n} + \mathbb{L}_{-2}^{bc}(z) e^{-i2 k_y} c_{n}  = 0 \; ,
\end{align}
\begin{align}
\label{1d_lieb_TBA_eq3}
  i  \frac{d c_{n}}{dz} &+ ( \Delta {\bf h}_{cb} \cdot {\bf A}_z +  \gamma_{\rm nl} |c_{n}|^2 ) c_{n} \\ \nonumber
 & + \mathbb{L}_2^{cb}(z) e^{i 2 k_y}  b_{n}  + \mathbb{L}_{-2}^{cb}(z) b_{n} = 0  \; .
\end{align}
}{We solve Eqs.~(\ref{1d_lieb_TBA_eq1})-(\ref{1d_lieb_TBA_eq3}) with {appropriate zero} boundary conditions to obtain all Lieb lattice edge modes shown below.}

\subsection{Kagome Tight-binding Approximation}
\label{sq_TBA}

Next we give the tight-binding approximation for the Kagome lattice. {As mentioned above, each} lattice site interacts with two of the other nearest 
 {site types} (e.g. $a$ sites interact with nearest $b$ and $c$ sites).
Similar to the Lieb case, we study a semi-infinite strip domain: zero boundary conditions on the left and right edges, and infinite boundary conditions along the $m$-direction. 
 {We} focus on two {types} of boundary conditions: ``pointy'' {(the analog of bearded for Lieb)} and ``straight'' (see Fig.~\ref{kagome_fig}). Any combination of these two boundary types can be accommodated.

Taking into account self and nearest neighbor interactions {and weak on-site nonlinearity} we arrive at the following system of equations describing this Floquet lattice {
\begin{align}
 \label{kagome_TBA_eq1}
 i \frac{d a_{mn}}{dz} +& ( \Delta {\bf h}_{ab} \cdot {\bf A}_z + \gamma_{\rm nl} |a_{mn}|^2 )a_{mn} \\ \nonumber
 +&    \mathbb{L}_{1}^{ab}(z)   b_{m+1,n+1} +  \mathbb{L}_{-1}^{ab}(z) b_{mn} \\ \nonumber
 + &\mathbb{L}_{3}^{ac}(z)  c_{m+2,n} +  \mathbb{L}_{-3}^{ac}(z) c_{mn}    = 0 \; , 
 \end{align}
 \begin{align}
 \label{kagome_TBA_eq2}
 i \frac{d b_{mn}}{dz} +& \gamma_{\rm nl} |b_{mn}|^2 b_{mn} \\ \nonumber
 +& \mathbb{L}_1^{ba}(z) a_{mn} + \mathbb{L}_{-1}^{ba}(z) a_{m-1,n-1}   \\ \nonumber
 + &    \mathbb{L}_2^{bc}(z) c_{mn} + \mathbb{L}_{-2}^{bc}(z) c_{m+1,n-1}  = 0 \; ,
  \end{align}
 \begin{align}
  \label{kagome_TBA_eq3}
  i \frac{d c_{mn}}{dz} +&  (  \Delta {\bf h}_{cb} \cdot {\bf A}_z +  \gamma_{\rm nl} |c_{mn}|^2 ) c_{mn} \\ \nonumber
 + & \mathbb{L}_3^{ca}(z) a_{mn} +  \mathbb{L}_{-3}^{ca}(z)  a_{m-2,n}    \\ \nonumber
 +& \mathbb{L}_2^{cb}(z)  b_{m-1,n+1} + \mathbb{L}_{-2}^{cb}(z)   b_{mn}  = 0 \; ,
\end{align}
where $\mathbb{L}_{\pm \ell}^{ij}(z) = \mathbb{L}(\pm {\bf v}_\ell - \Delta {\bf h}_{ij}(z)).$} The coefficients  {$\mathbb{L}$ and $\gamma_{\rm nl}$ are defined in Appendix \ref{TBA_coefficients}. 

To find edge modes localized along the left or right boundaries, we consider solutions of the form {
\begin{align}
\nonumber
& a_{mn}(z) = a_{n}(k_y; z) e^{i m k_y } \; , \\
& b_{mn}(z) = b_{n}(k_y; z) e^{i m k_y } \; , \\ \nonumber
& c_{mn}(z) = b_{n}(k_y; z) e^{i m k_y } \; ,
\end{align}
to obtain the reduced system of equations {
\begin{align}
 \label{1d_kagome_TBA_eq1}
 i \frac{d a_{n}}{dz} +& ( \Delta {\bf h}_{ab} \cdot {\bf A}_z + \gamma_{\rm nl} |a_{n}|^2 )a_{n} \\ \nonumber
 +&  \mathbb{L}_{1}^{ab}(z)  e^{i k_y}  b_{n+1} +  \mathbb{L}_{-1}^{ab}(z) b_{n} \\ \nonumber  
 + & \mathbb{L}_{3}^{ac}(z) e^{i 2k_y}   c_{n} +  \mathbb{L}_{-3}^{ac}(z) c_{n} = 0 \; ,
 \end{align}
 \begin{align}
 \label{1d_kagome_TBA_eq2}
 i \frac{d b_{n}}{dz} + & \gamma_{\rm nl} |b_{n}|^2 b_{n} \\ \nonumber
 +& \mathbb{L}_1^{ba}(z) a_{n} + \mathbb{L}_{-1}^{ba}(z)  e^{-ik_y}  a_{n-1}  \\ \nonumber
 + & \mathbb{L}_2^{bc}(z) c_{n} + \mathbb{L}_{-2}^{bc}(z) e^{ik_y} c_{n-1}   = 0 \; ,
  \end{align}
 \begin{align}
  \label{1d_kagome_TBA_eq3}
  i \frac{d c_{n}}{dz} +&  (  \Delta {\bf h}_{cb} \cdot {\bf A}_z +  \gamma_{\rm nl} |c_{n}|^2 ) c_{n} \\ \nonumber
 + &  \mathbb{L}_3^{ca}(z) a_{n} +  \mathbb{L}_{-3}^{ca}(z)  e^{-i 2k_y} a_{n} \\ \nonumber  
 +&  \mathbb{L}_2^{cb}(z) e^{- i k_y}  b_{n+1} + \mathbb{L}_{-2}^{cb}(z)   b_{n} = 0 \; . 
\end{align}
}We solve Eqs.~(\ref{1d_kagome_TBA_eq1})-(\ref{1d_kagome_TBA_eq3}) with appropriate spatial boundary conditions  to find all Kagome edge modes shown below.

\section{Linear Floquet Bands and Edge State Dynamics}
\label{linear_theory}

We now compute linear ($\gamma_{\rm nl} = 0$) edge states for the tight-binding systems {given} 
above. To accomplish this we compute the principal fundamental matrix for the reduced systems (\ref{1d_lieb_TBA_eq1})-(\ref{1d_lieb_TBA_eq3}) or (\ref{1d_kagome_TBA_eq1})-(\ref{1d_kagome_TBA_eq3}) after one period in $z$. From this we obtain the monodromy matrix at $z = T$ (where $T$ is the helix pitch or period: $T=2\pi/\Omega$, and $\Omega \equiv \Omega_b$). The eigenvalues of the monodromy matrix are the Floquet (characteristic) multipliers  $\lambda(k_y)$. The Floquet exponents are computed, up to an additive constant, by
\begin{equation}
\label{floquet_exponent}
\alpha(k_y) =  \frac{i \ln [\lambda(k_y)]}{T} + \frac{2 \pi l}{T} \; , ~~ l \in \mathbb{Z} \; .
\end{equation}

{Since the range of possible lattice rotation patterns is very large} we focus our attention on driving patterns which have been shown to demonstrate interesting 
 {band} structures in other lattice systems. Previous examples include in-phase rotation in honeycomb lattices \cite{RechtsSegev2013}, or $\pi$-phase offset among the sublattices {that} generate Weyl type-II points \cite{Noh2017}.
In particular we concentrate on the following driving patterns:
\begin{itemize}
\item same (in-phase) rotation
\begin{equation}
\label{same_rot_func}
{\bf h}_{j}(z) = \eta \left( \cos\left( \Omega  z  \right) , \sin\left( \Omega  z \right) \right) \; ,
\end{equation}
\item different radii, in-phase 
{\begin{equation}
\label{diff_rad_rot_func}
 {\bf h}_{j}(z) = R_j \eta \left( \cos\left( \Omega  z  \right) , \sin\left( \Omega  z \right) \right) \; , ~~ R_j \le 1
\end{equation}}
\item $\pi$-phase offset
\begin{equation}
\label{pi_phase_rot_func}
{\bf h}_{j}(z) =  \eta \left( \cos\left( \Omega z + \chi_j \right) , \sin\left( \Omega  z + \chi_j \right) \right) \; , ~~ \chi_j = 0,\pi
\end{equation}
\item counter rotation 
\begin{equation}
\label{counter_rot_func}
{\bf h}_{j}(z) = \eta \left( \cos\left( \Omega  z  \right) , r_j \sin\left( \Omega  z \right) \right)  \; , ~~ r_j = \pm1
\end{equation}
\item different frequency
\begin{equation}
\label{diff_freq_rot_func}
{\bf h}_{j}(z) = \eta \left( \cos\left(  l_j \Omega  z  \right) , \sin\left(  l_j \Omega  z \right) \right) \; , ~~ l_j \in \mathbb{N} \; ,
\end{equation}
\item quasi one-dimensional motion
\begin{align}
\label{1d_motion}
& {\bf h}_{j}(z) = \eta \left(p_j  \cos\left( \Omega z  \right) , q_j \sin\left( \Omega  z \right) \right) \; ,\\ \nonumber
& p_j = 0~{\rm or}~1 \; , ~~~~   p_j + q_j = 1 \; ,
\end{align}
\end{itemize}
where  $j = a,b,c$, and $\eta \equiv \eta_b$. We point out that each of 
the driving functions above have period $T$ i.e. ${\bf h}_j(z + T) = {\bf h}_j(z),$ where {$T = 2\pi / \Omega$} and do not affect the translation symmetry (periodicity in ${\bf r}$). The reason is that each individual sublattice maintains a periodic spatial structure (square for Lieb, triangular for Kagome), and since the potential (\ref{general_potential}) is simply the sum of these sublattices, it too must preserve translation invariance.

For all examples considered in this paper we use lattice depth $V_0^2 = 45$ and angular frequency $\Omega = 2\pi/1.5;$ this corresponds a helix pitch of $T = 1.5.$
These parameters are taken to model the experimental parameters reported in \cite{RechtsSegev2013}.

\subsection{Lieb Floquet bands}

In this section we compute the Floquet bands {and location of edge modes} for the Lieb lattice. These bands, defined in Eq.~(\ref{floquet_exponent}), are calculated for the linear {$(\gamma_{\rm nl} = 0)$} system of equations given in Eqs.~(\ref{1d_lieb_TBA_eq1})-(\ref{1d_lieb_TBA_eq3}). The band structures consist of bulk or extended modes (indicated by solid {black} regions) and {gapless} 
edge or localized modes (indicated by curves, highlighted {with} color). 

{In the absence of helical driving ($\eta = 0$) the Lieb dispersion bands are known to possess a flat band that spans the Brillouin zone {cf.}~\cite{GuzmanSilva2014}. Regardless of the left/right boundary conditions (e.g. bearded/bearded, bearded/straight, or straight/straight) the band diagram resembles that shown in Fig.~\ref{lieb_stationary_band_color}. {{Though not visible, within the green colored} flat band region there exist both bulk modes {\it and} localized edge modes.} 
 Flat {band}  {edge modes are found to exist in the neighborhood of both edge types.} 

\begin{figure} [ht]
\includegraphics[scale=.5]{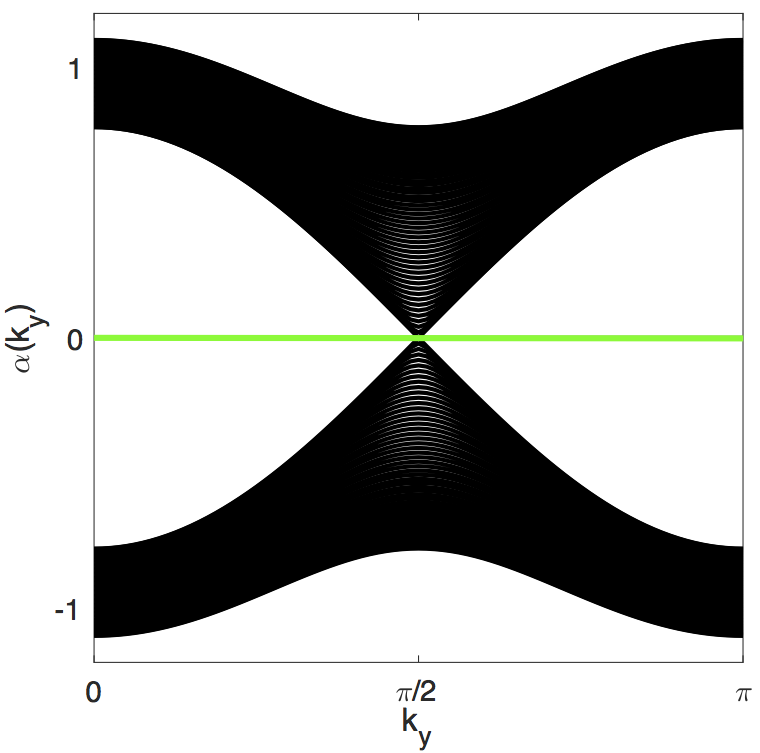}
\caption{(Color online) Lieb Floquet bands with no lattice driving. Green curve indicates that flat band modes are located {on/near} both edges. The parameters used are: $\eta = 0, \sigma_x = \sigma_y = 0.3.$}
\label{lieb_stationary_band_color}
\end{figure}

The first case {with driving} is that of same rotation (\ref{same_rot_func}) among the waveguides. Several typical Floquet bands are shown in Fig.~\ref{linear_lieb_bands_same} for different boundary conditions. 
Each case contains traveling ($\alpha'(k_y) \not= 0$)  and stationary ($\alpha'(k_y) = 0$) modes; {at least} one of each type on both {sides} 
 (left and right). { {Again,} since it is not visible, we also point out that at the flat band there are both bulk and edge modes.}
 
Each boundary type is found to exhibit a distinct signature in its band structure. 
For instance, the Floquet bands on the straight edge possess a steep slope (large group velocity), particularly near $k_y = \pi/2$, whereas the bearded edge modes have a shallow slope (smaller group velocity) throughout the Brillouin zone. As a consequence, for this rotation pattern there {are} fast edges (straight) and slow edges (bearded).

{To establish these are indeed topologically protected edge modes we also calculate the Chern number for each bulk band. The definition and computation of the Chern number is described in Appendix \ref{chern_section}. {The Bulk-edge correspondence states that the Chern numbers in both the bulk and edge problems are the same \cite{Hatsugai1993}. Furthermore, the Chern number is equal to the net number of chiral edge modes above the band minus the number below the band \cite{Lu2014,Rudner2013}. A corollary of this is that the Chern number is independent of the boundary used; hence for a set of bulk bands with different Chern numbers (see below) there will be a gapless edge state; even in the presence of a lattice defect.} The nontrivial Chern number in the top and bottom spectral bands {of Fig.~\ref{linear_lieb_bands_same}} indicates the presence of topological edge states. Since the flat band has the same number of chiral modes above and below, 
 it's Chern number is zero. 
This figure also highlights that the spectral bulk bands are {\it unaffected} by the boundary conditions.}



\begin{figure} [ht]
\includegraphics[scale=.37]{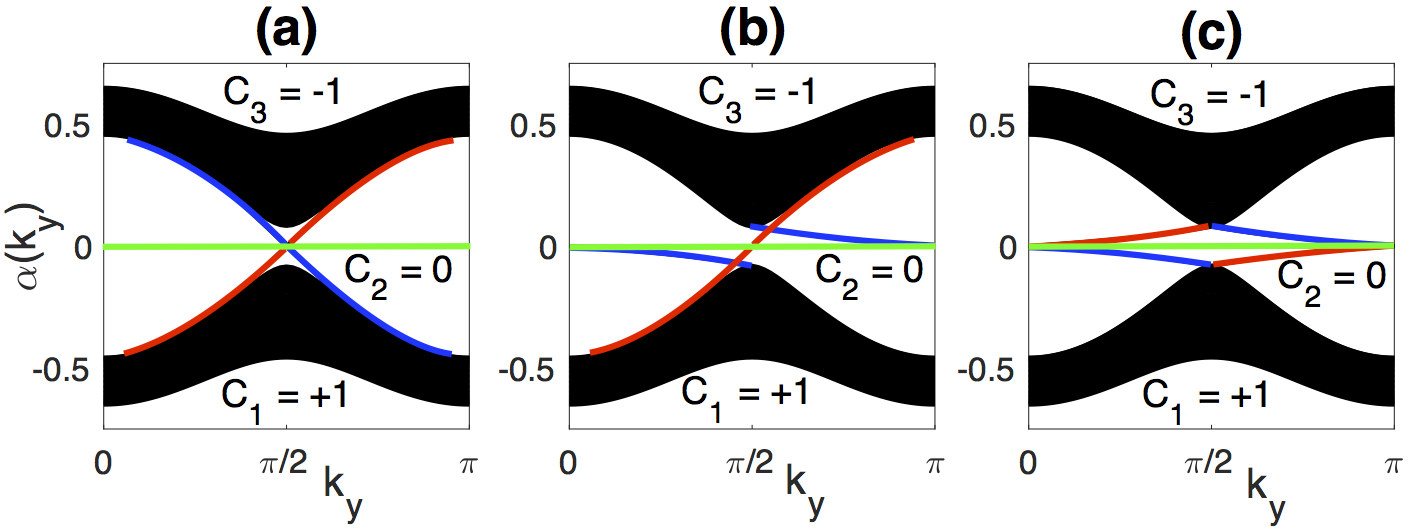}
\caption{(Color online) Lieb Floquet bands for same phase lattice rotation (\ref{same_rot_func}) with different boundary conditions. The boundary condition on the left [right] edge is: (a) straight [straight], (b) bearded [straight], and (c) bearded [bearded]. Red curves indicate edge modes on right edge, blue curves denote left edge modes, and the green curves designate flat band modes on both edges. The parameters used are: $\eta = 2/3, \sigma_x = \sigma_y = 0.3.$ {Chern numbers for each bulk band are included.}}
\label{linear_lieb_bands_same}
\end{figure}

For all remaining band calculations we use {the} boundary combination shown in Fig.~\ref{lieb_fig}, namely bearded on the left and straight on the right.
 {The} next case considered is that of $\pi$-offset rotation (\ref{pi_phase_rot_func}). 
The corresponding  Floquet bands are displayed in Fig.~\ref{linear_lieb_bands_pi} for three different values of $\eta$ {i.e. {radii}}. {Each case contains a set of topologically protected edge states residing in the central gaps.} One significant difference between these bands and those found in Fig.~\ref{linear_lieb_bands_same}(b) 
is the absence of any flat band modes. 
In addition, at a certain radius threshold [see Fig.~\ref{linear_lieb_bands_pi}(b)] the gap between different branches of the Floquet exponent (\ref{floquet_exponent}) closes {and the system undergoes a topological transition}. {Since it is not possible to distinguish between different bulk bands at this transition point, we do not calculate Chern numbers in Fig.~\ref{linear_lieb_bands_pi}(b).}



Increasing the helix radius beyond this threshold spawns an entirely new family of edge states located near the {edge of the Floquet region:} {$\alpha_{\rm edge} = \pm \pi/T$}. {Since {now} there is one chiral mode above and below each bulk band, the Chern number is zero.}
{Finally, we point out that the bearded edge modes shown in Fig.~\ref{linear_lieb_bands_pi} move considerably faster than those in Fig.~\ref{linear_lieb_bands_same}(b). Hence it is possible, (by changing the lattice rotation pattern) to support faster edge mode propagation along a bearded edge.}



\begin{figure} [ht]
\includegraphics[scale=.38]{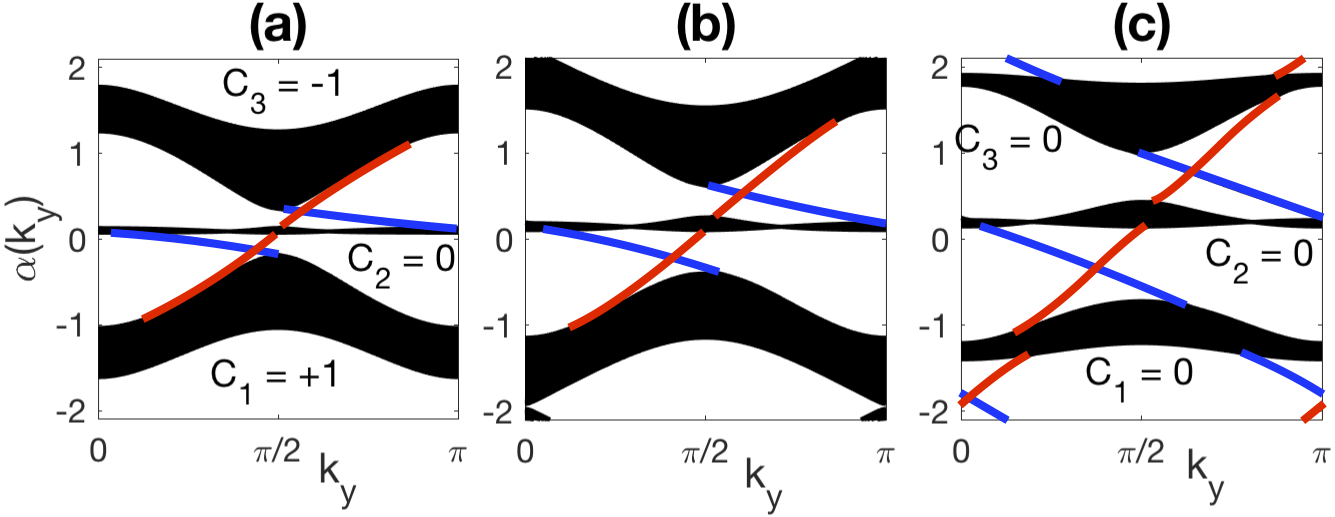}
\caption{(Color online) Lieb Floquet bands for $\pi$-offset rotation (\ref{pi_phase_rot_func}).
The boundary condition on the left [right] edge is bearded [straight]. Red curves indicate edge modes on right edge, blue curves denote left edge modes. The parameters used are: (a) $\eta =1.4/15$, (b) $\eta =1.8/15$, and (c) $\eta = 2.2/15$ with {$ \chi_a = \chi_c = \pi, \chi_b = 0, \sigma_x = \sigma_y = 0.3.$}}
\label{linear_lieb_bands_pi}
\end{figure}

Many other lattice rotation patterns are also found to support {topologically protected} 
edge states. In Fig.~\ref{linear_lieb_bands_multi} we show Floquet bands for various driving patterns on the Lieb lattice. The dispersion curves corresponding to same rotation driving (\ref{same_rot_func}) with elliptical ($\sigma_x \not= \sigma_y$), rather than circular, waveguides {are} shown in  Figs.~\ref{linear_lieb_bands_multi}(a) and \ref{linear_lieb_bands_multi}(b). In the former case ($\sigma_x < \sigma_y$) the major axis of the ellipse is parallel to the $m$-direction, while in the latter scenario ($\sigma_x > \sigma_y$) the major axis is parallel to the $n$-direction. Doing this is found to either squeeze [see Fig.~\ref{linear_lieb_bands_multi}(a)] or stretch [see Fig.~\ref{linear_lieb_bands_multi}(b)] the width of the {upper and lower} bulk bands. 
We find that this lattice rotation arrangement does support {both extended (bulk) and localized (edge) flat band modes, both at $\alpha(k_y) = 0$.} 

Next we consider when the sublattices are rotating in-phase with each other, but with different radii (\ref{diff_rad_rot_func}). 
Examining the corresponding band structure in Fig.~\ref{linear_lieb_bands_multi}(c) we observe the presence of unidirectional modes and the absence of any flat band states. The next set of bands {[see Fig.~\ref{linear_lieb_bands_multi}(d)]} correspond to different frequency (\ref{diff_freq_rot_func}) among the sublattices. In particular we examine when the $a$ and $c$ lattice sites rotate at twice the frequency of the $b$ sites. 
Overall, the band {and Chern} structure resembles {that of the} same rotation case shown in Fig.~\ref{linear_lieb_bands_same}(b) without, however, the presence of any flat band modes. The bands corresponding to counter rotation (\ref{counter_rot_func}) are shown in Fig.~\ref{linear_lieb_bands_multi}(e). 
 {For these parameters} the $b$ sites are rotating in a counter-clockwise fashion while the $a$ and $c$ sites move in the clockwise direction. No localized modes are found 
for {these} {parameters}{, and only trivial Chern numbers are found.} 

The final case is that of quasi one-dimensional motion (\ref{1d_motion}) where each sublattice moves in only one {direction}, {either the $m$ or $n$ direction}. We consider a scenario in which the $a$ and $c$ sites oscillate only in the $m$-direction ($p_a = p_c = 0, q_a = q_c = 1$), while the $b$ sites oscillate only in the $n$-direction ($p_b = 1, q_b = 0$). The corresponding bands are shown in Fig.~\ref{linear_lieb_bands_multi}(f). For this relatively simple lattice motion we identify the presence unidirectional traveling edge modes. Looking closer we note that these modes travel in an orientation opposite those found in previous cases {i.e. they possess opposite chirality}. In contrast to the previous cases, here the straight (right) edge modes move in the negative direction ($\alpha'(k_y) < 0$), while {bearded} (left) edge modes travel in the positive direction ($\alpha'(k_y) > 0$). {This is reflected in the Chern numbers of the upper and lower bulk bands which have opposite signs compared to the previous cases.} No flat band modes are found.




\begin{figure} [ht]
\centering
\includegraphics[scale=.6]{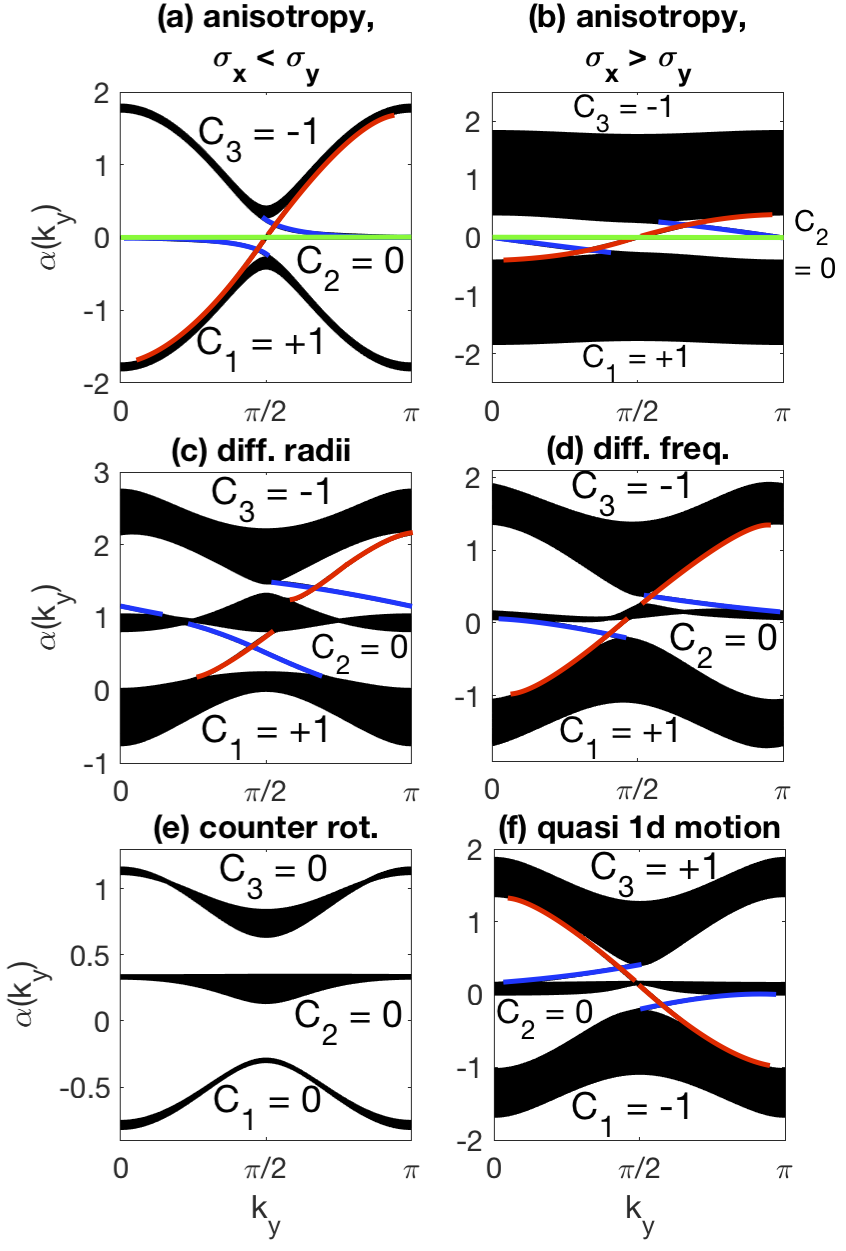}
\caption{(Color online) Lieb Floquet bands for various rotation patterns. The boundary condition on the left [right] edge is bearded [straight]. Red curves indicate edge modes on right edge, blue curves denote left edge modes, and the green curves designate flat band modes on both edges. The parameters used are: {(a-b) $\eta =2/3$, (c) $\eta = 2/3, R_a = R_c = 3/5, R_b = 1$, (d) $\eta = 2.2/15, l_a = l_c = 2, l_b = 1$, (e) $\eta =3/15, r_a = r_c = -1, r_b = 1$, (f) $\eta = 3/15, p_a = p_c = 0, p_b = 1$;  with $\sigma_x = \sigma_y = 0.3$ in all except (a) $\sigma_x=0.3,\sigma_y=0.5$ and (b) $\sigma_x=0.5,\sigma_y=0.3$.} {Chern numbers for each bulk band are included.}}
\label{linear_lieb_bands_multi}
\end{figure}

\subsection{Lieb edge mode dynamics}

In this section we present {mode} evolutions for {some} the edge states found above.
Specifically, we integrate the full Lieb tight-binding {coupled mode} system (\ref{lieb_TBA_eq1})-(\ref{lieb_TBA_eq3}) for an edge mode with a localized envelope in $m$. 
The initial conditions {taken are} { 
\begin{align}
\nonumber
&a_{mn}(0)  =   {\rm sech}\left(  \mu m  \right) a_{n}(k_y) e^{i 2 m k_y } \; , \\ \label{linear_Lieb_IC}
&b_{mn}(0) = {\rm sech}\left( \mu m  \right) b_{n}(k_y) e^{i 2m  k_y } \; , \\ \nonumber
&c_{mn}(0) = {\rm sech}\left( \mu m  \right) c_{n}(k_y) e^{i 2m k_y } \; ,
\end{align}
}{using a typical value} {$\mu = 0.1$}. The {exponentially decaying} edge eigenmodes {$a_{n}(k_y),b_{n}(k_y),$ and $c_{n}(k_y)$} are obtained directly from solving system (\ref{1d_lieb_TBA_eq1})-(\ref{1d_lieb_TBA_eq3}) {at a chosen $k_y$}.
 {In all cases} the eigenmode two-norm is fixed to one i.e. {$\sum_n \left( |a_{n}|^2 + |b_{n}|^2 + |c_{n}|^2 \right)= 1.$} We take periodic boundary conditions in $m$ (top/bottom edges) and {bearded}-straight zero boundary conditions in $n$. The left-most lattice site is located at {$n = 0$}, while the right-most site is $n = N$, where $N$ is taken to be large $\sim O(100)$.
 The system is integrated using a fourth-order Runge-Kutta method.
 
The $z$-dynamics for several edge mode profiles are presented in Fig.~\ref{lieb_linear_evolve}. For simplicity of presentation, we only show the most dominant (largest magnitude) sublattice mode {(e.g. the $c$-mode tends to be considerably larger than the $a$-mode near the straight side)}. The evolutions shown in Figs.~\ref{lieb_linear_evolve}(a-d) correspond to same rotation among the three sublattices, the Floquet bands of which are shown in Fig.~\ref{linear_lieb_bands_same}(b). The two {topological} {traveling} 
modes ($\alpha'(k_y) \not= 0$) are observed to propagate with constant velocity in either the negative (on the left side)  [see Fig.~\ref{lieb_linear_evolve}(a)] or positive (on the right side) direction [see Fig.~\ref{lieb_linear_evolve}(b)]. As expected, the mode on {the} bearded edge travels considerably slower than the straight edge mode. {In order to get a well-localized {edge} mode the value of {$\alpha(k_y)$} must be chosen well-separated from the bulk bands. This is why the same rotation mode shown in Fig.~\ref{lieb_linear_evolve}(a) has a different {mode number} $k_y$ than the others. When an edge mode with {a} {corresponding} Floquet exponent {located} near {a} bulk band is used the mode is found to {\it not} maintain its {well-}localized structure over long distances and {will} 
{\it not} propagate through lattice defects. {In a sense such modes can be considered to be quasi-bulk bands.} 
This is discussed in Sec.~\ref{defect}.  

\begin{figure} [ht]
\centering
\includegraphics[scale=.68]{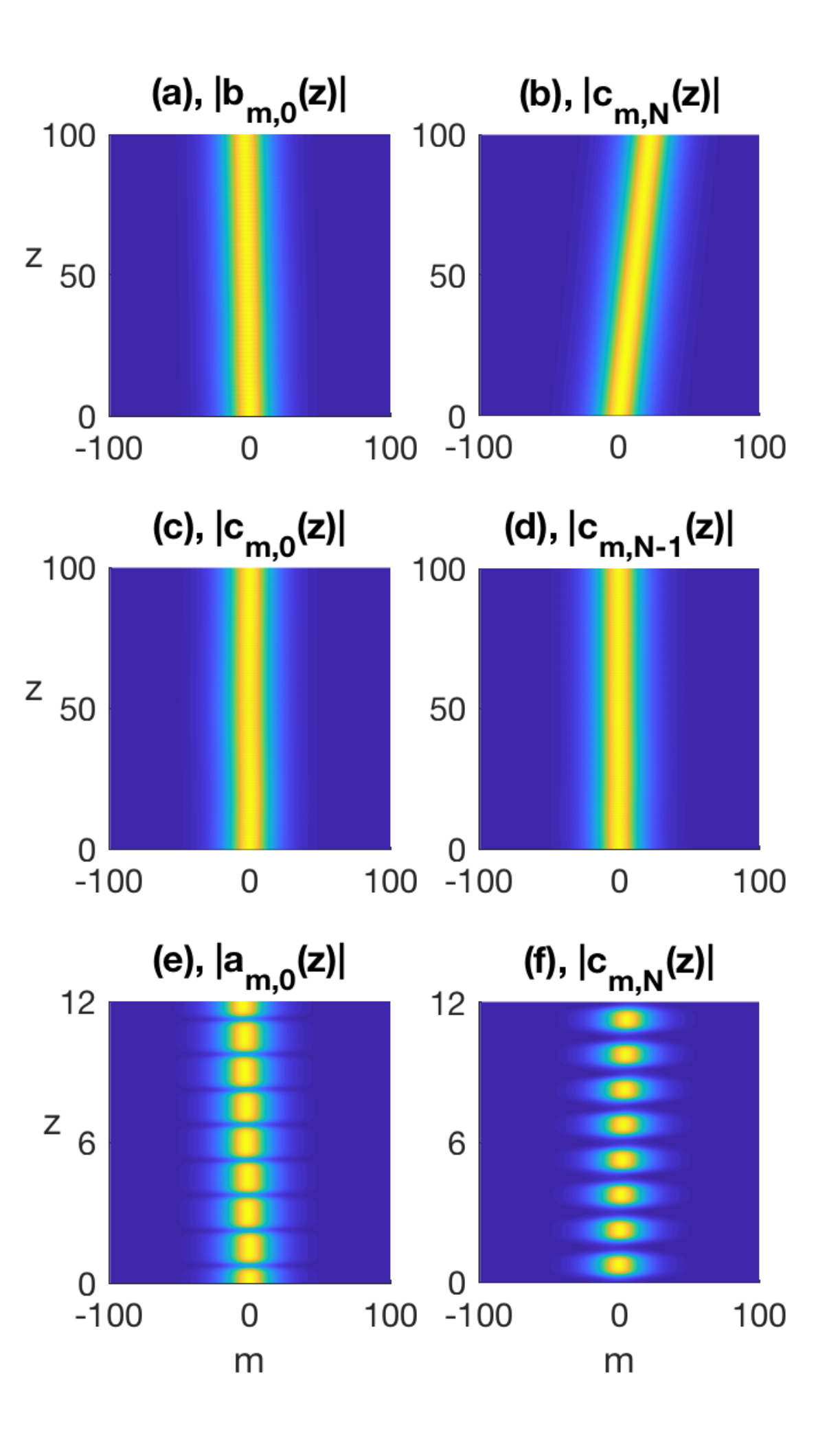}
\caption{Evolution of edge profile in the Lieb lattice. The edge modes shown in panels (a-d) correspond to the Floquet bands in Fig.~\ref{linear_lieb_bands_same}(b) {at: $(k_y, \alpha) =$ (a) $(1.8,. 0628) $, (b) $(1.65,.0362)$, (c-d) $(1.65,0)$.} The edge states in panels (e-f) correspond to the bands shown in  Fig.~\ref{linear_lieb_bands_pi}(c) at {$(k_y, \alpha) =$ (e) $(.2,-1.94)$ and (f) $(.2,-1.68)$.}
}\label{lieb_linear_evolve}
\end{figure}

Next we evolve the stationary flat band states in Fig.~\ref{linear_lieb_bands_same}(b). When solving for the Floquet exponent on the straight edge the {numerical} algorithm we use produces two stationary modes{:} one with {$|\alpha(1.65)| = O(10^{-4})$} and another at {$|\alpha(1.65)| = O(10^{-11})$}. In Fig.~\ref{lieb_linear_evolve} we only consider the {latter} mode whose 
 {magnitude} is smaller. 
 {We note that} the magnitude of these flat band modes is found to peak not along the boundary, but instead at an interior column {(at {$n = N-1$} on the right)}. The flat band evolutions are shown in Figs.~\ref{lieb_linear_evolve}(c-d). As expected, these modes do not move from their initial position.


The final set of evolutions we present is that of $\pi$-offset rotation. We omit the edge mode dynamics for the bands shown in Fig.~\ref{linear_lieb_bands_multi}, but note that many similar evolution patterns are observed in those cases {as well}.
We focus on the $\pi$-offset modes near the Floquet edge in Fig.~\ref{linear_lieb_bands_pi}(c).
The corresponding mode evolutions are displayed in Figs.~\ref{lieb_linear_evolve}(e-f). The {mode profiles} on both sides {are} observed to oscillate with the same period as the {lattice driving} ($T = 1.5$). Moreover along the bearded edge [see Fig.~\ref{lieb_linear_evolve}(e)] energy fluctuates back-and-forth between the {$a/ c$ sites} (in-phase with each other) and the $b$-sites {(out-of-phase)}. {{If the} overall beam (\ref{ansatz_define}) is {wide 
this oscillatory behavior may be difficult to detect.} However, along the {$n=0$} column there should be an {noticeable} oscillation in the beam intensity.}
In Fig.~\ref{lieb_linear_evolve}(f) 
 {the straight edge mode shown, {$c_{m,N}$},} has a similar, but out-of-phase, evolution pattern with the $b$-sites; meanwhile the $a$ lattice modes are relatively small in comparison.


\subsection{Kagome Floquet bands}

In this section we {compute}  Floquet bands for the Kagome lattice.
The dispersion curves are computed in the same way as the Lieb bands in the previous section, via  the Floquet exponents (\ref{floquet_exponent}), for the rotation patterns listed in Eqs.~(\ref{same_rot_func})-(\ref{1d_motion}). 
All bands are computed from the one-dimensional tight-binding system given in Eqs.~(\ref{1d_kagome_TBA_eq1})-(\ref{1d_kagome_TBA_eq3}). 

{Before considering {a} driven case we first examine the dispersion bands {for stationary waveguides} ($\eta = 0$). 
The Floquet bands for different boundary conditions are displayed in Fig.~\ref{kagome_no_rot_compare}, where both traveling and non-traveling edge states are observed. Regardless of the edge type we find a flat top band. By introducing a pointy edge (see Fig.~\ref{kagome_fig}) we observe a ``vine'' type family of {edge} modes which {do} not span {the} gap.  
 {To our knowledge flat band {{\it edge}} modes have not previously been}  considered in the context of a {kagome} photonic lattice waveguide. Interestingly, below we do not find any flat bands like these in the presence of periodic driving (unlike the Lieb lattice above).} 


\begin{figure} [ht]
\includegraphics[scale=.47]{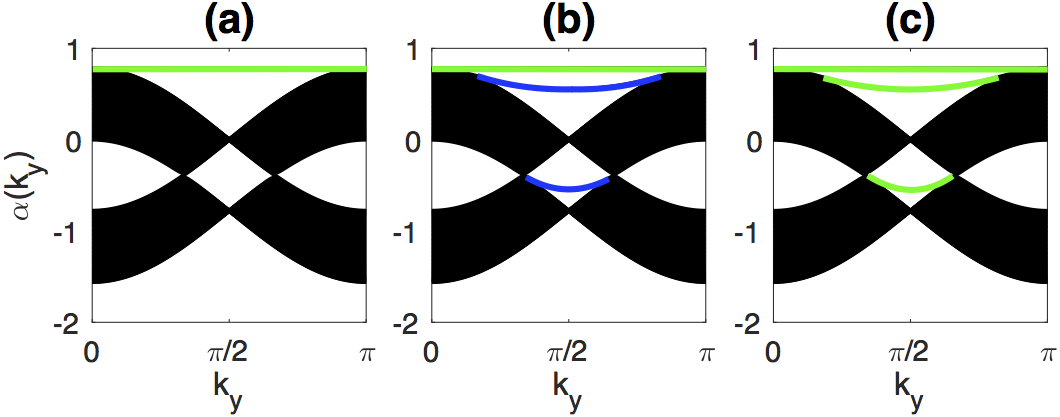}
\caption{(Color online) Kagome Floquet bands in the absence of rotation with different boundary conditions. The boundary condition on the left [right] edge is: (a) straight [straight], (b) pointy [straight], and (c) pointy [pointy]. Blue curves denote left edge modes and the green curves designate modes on both edges. The parameters used are: {$\eta = 0, \sigma_x = \sigma_y = 0.3.$}}
\label{kagome_no_rot_compare}
\end{figure}

The first {driven} case to consider is that of same rotation among all sublattices (\ref{same_rot_func}). The corresponding band diagrams  are shown in Fig.~\ref{kagome_same_rot_diff_BC} for three different boundary combinations. {Also included are the corresponding Chern numbers of the bulk bands. Here and below we highlight with color {only} the (topological) gapless edge modes.} 
Similar to the Lieb lattice above, here we observe that different boundary conditions carry distinguishable {edge} band structure. Both edge types possess a {topological} edge mode in the central gap ({$-0.5 \le \alpha(k_y) \le 0$}). {These gapless modes 
have sign-definite group velocity (unidirectionality) throughout the gap.} 
For the straight edge there are edge modes near {$\alpha = 0.5$ that span a very small gap between the middle at top bulk bands. {To see that the mode actually crosses the gap one must zoom in very close.}
On the other hand, when a pointy edge is {introduced} 
we observe the} ``vine'' type curve in the upper gap ($0.3 \le \alpha(k_y) \le 0.4$). {In either case the bands in the upper gap do span the entire gap ({and so they are highlighted with color}).} We point out that the pointy ``vine'' edge curves are not slope-definite, {in other words} {depending on $k_y $ the} group velocity may be either positive ($\alpha'(k_y) > 0$) or negative ($\alpha'(k_y) < 0$). {{Also,} while the top bulk band looks flat, it is not. This is different than the Lieb lattice in Fig.~\ref{linear_lieb_bands_same} {where there is a flat band at $\alpha = 0$.}}



\begin{figure} [ht]
\includegraphics[scale=.46]{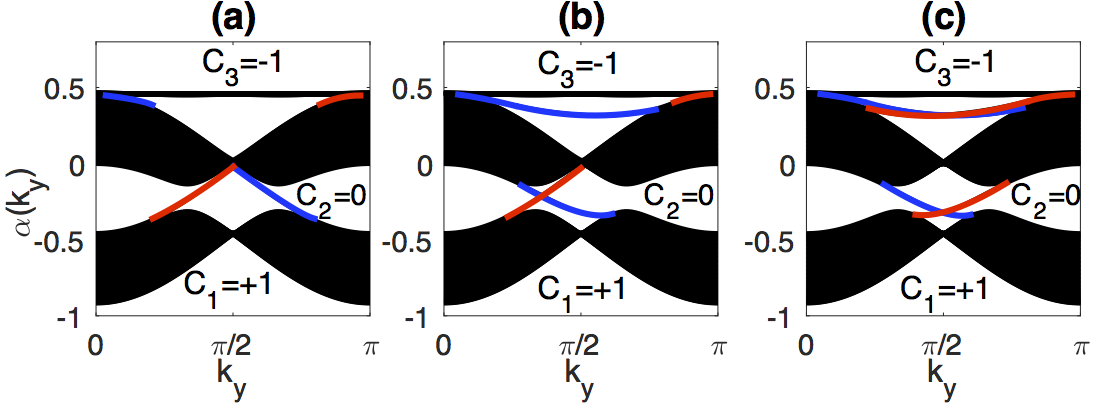}
\caption{(Color online) Kagome Floquet bands for same phase lattice rotation (\ref{same_rot_func}) with different boundary conditions. The boundary condition on the left [right] edge is: (a) straight [straight], (b) pointy [straight], and (c) pointy [pointy]. Red curves indicate edge modes on right edge, blue curves denote left edge modes. The parameters used are: {$\eta = 2/3, \sigma_x = \sigma_y = 0.3.$} {Chern numbers for each bulk band are included.}}
\label{kagome_same_rot_diff_BC}
\end{figure}

For {the} remaining {Kagome} band diagrams we consider pointy-straight boundary conditions, similar to those in Fig.~\ref{kagome_fig}.
When the individual waveguides are elliptical in shape (major axis parallel to the $n$-direction, $\sigma_x > \sigma_y$) and all sublattices are rotating in-phase with each other (\ref{same_rot_func}), we find the band diagram shown in Fig.~\ref{kagome_multi_band}(a). {While these bands have the same Chern structure as those found in Fig.~\ref{kagome_same_rot_diff_BC}(b),} this band structure bears little resemblance to the isotropic {case}. 
The pointy edge (blue) bands are observed to have slopes of positive {or} negative sign {indicating that the system admits modes that may travel in either direction. {We categorize this mode as topological since it spans the gap and corresponds to a nontrivial Chern number; even though it is not unidirectional throughout the entire gap. Moreover, If we evolve a mode frequency at a frequency $\alpha$ that does not support any other bulk or edge modes, we do observe unidirectional motion.}  {Additionally, there is a set of topological modes located on the straight edge that span the central gaps.} 






We next examine different radii among the sublattices (\ref{diff_rad_rot_func}). In particular we consider when the $a$ and $c$ sublattices have smaller radii ($R_a = R_c = 0.6$) compared to the $b$ lattice sites ($R_b = 1$). The corresponding band diagram is shown in Fig.~\ref{kagome_multi_band}(b). Numerous {gapless} modes on both edges are found{; all in agreement with the Chern numbers.} One unusual family of edge states is located on the straight edge and {have} sign-indefinite group velocity, {yet still span the gap}  [see minimum point near $(k_y, \alpha) = (1.7,1.4)$]. {{Indeed}, when {an envelope was formed nearby this point and} 
propagated into a defect barrier it did backscatter.} 





{Th}e Floquet bands corresponding to different frequency among the sublattices {--see: Eq.~(\ref{diff_freq_rot_func}})-- are shown in Fig.~\ref{kagome_multi_band}(c). Here we take the $a$ and $c$ sublattices to oscillate twice as fast as the $b$ sites. 
 {Both {interior} gaps contain topological modes that span their respective gaps. In the upper gap there is an additional non-topological band {(colored in black)} which does not span the gap, and hence does not affect the Chern invariant. The {pointy edge band in the upper gap (colored in blue)} has a small negative slope.}}


\begin{figure} [ht]
\includegraphics[scale=.67]{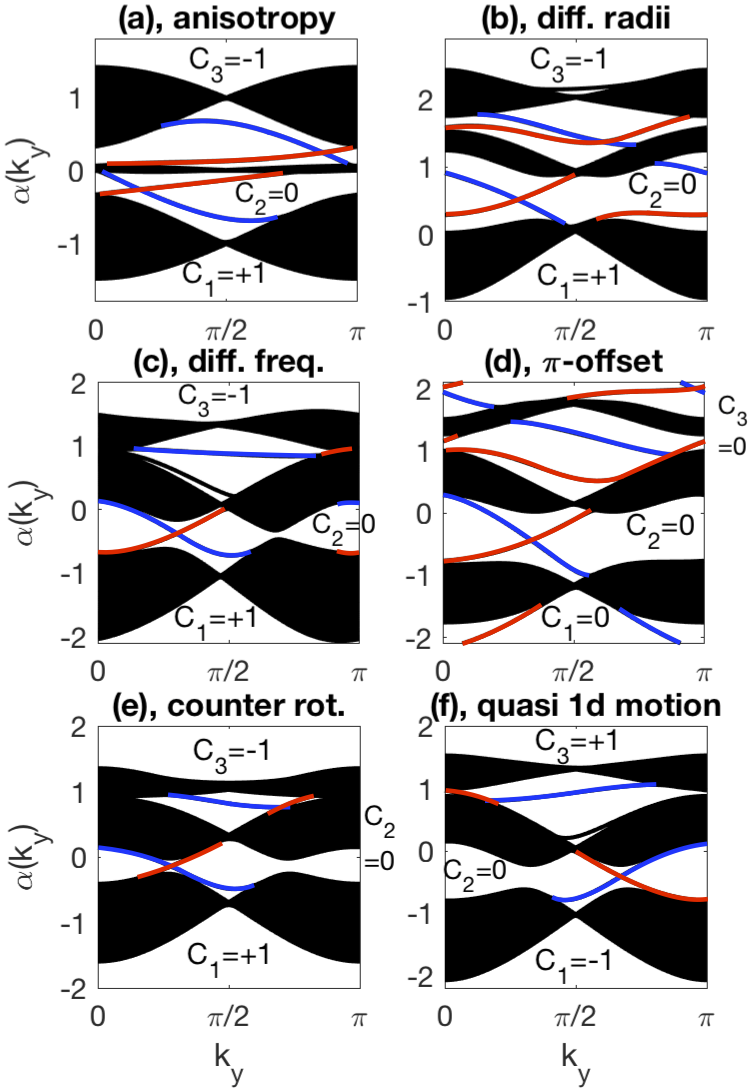}
\caption{(Color online) Kagome Floquet bands for various rotation patterns. The boundary conditions on the left [right] edge is pointy [straight]. Red curves indicate {gapless} edge modes on right edge, blue curves denote {gapless}  left edge modes. The parameters used are: {(a) $\eta=2/3$,  {(b) $\eta = 2/3, R_a = R_c = 3/5, R_b = 1$,} (c) $\eta = 2.2/15, l_a = l_c = 2, l_b = 1$, (d) $\eta = 2.2/15, \chi_a = \chi_c = \pi , \chi_b = 0$,  (e) $\eta = 1/5, r_a = r_c = -1, r_b = 1$,  (f) $\eta =1/5, p_a = p_c = 0, p_b = 1$; $\sigma_x = \sigma_y = 0.3.$ in all except (a) where $\sigma_x=0.5,\sigma_y=0.2$.} {Chern numbers for each bulk band are included.}}
\label{kagome_multi_band}
\end{figure}

When a $\pi$-phase offset (\ref{pi_phase_rot_func}) is introduced between the sublattices  
a threshold phenomena, similar to that observed in Fig.~\ref{linear_lieb_bands_pi} for the Lieb lattice, is found to occur. Increasing the lattice driving (helix {radius}) we observe the gap between adjacent Floquet exponent bands (\ref{floquet_exponent}) close at some threshold value $\eta_T > 0$ and then reopen with a new family of edge states for $\eta > \eta_T$. In Fig.~\ref{kagome_multi_band}(d) we show the bands above this threshold{; here we see} {numerous {gapless modes} on both edges}. {Since the net {number} of chiral modes above and below the spectral bands are the same, the resulting Chern number is zero for each.} 



Counter rotation among the sublattices (\ref{counter_rot_func}), specifically  when the $a$ and $c$ lattice sites have the opposite orientation to that of the $b$ sites, 
is considered next. The corresponding bands are shown in \ref{kagome_multi_band}(e). In contrast to the Lieb lattice above [see Fig.~\ref{linear_lieb_bands_multi}(e)], here we {\it do} find some edge waves. {Both Floquet bands in the upper gap have slope that is sign-definite, meanwhile the pointy (left) edge mode in the {lower} gap has a small strip {where there are both negative and positive} directional modes.} 

{The} final scenario we investigate is that of the quasi one-dimensional rotation pattern (\ref{1d_motion}). A band diagram for this arrangement is shown in Fig.~\ref{kagome_multi_band}(f). The $a$ and $c$ lattice sites only move in the $m$-direction ($p_a = p_c = 0$), while the $b$ sites only move in the $n$-direction ($q_b = 0$). Relative to several previous cases, these edge modes {tend to move} in a clockwise (as opposed to counter-clockwise) fashion. {As a result, the gap modes have opposite chirality and the corresponding Chern number have opposite sign from those found in the previous rotation patterns.} 





It is worth {mentioning some of the} similarities between the Lieb and Kagome lattices. 
 {Gapless} edge states are found for same sublattice rotation (\ref{same_rot_func}) {in both cases}. 
 Introducing a $\pi$-offset (\ref{pi_phase_rot_func}) among the sublattices introduces a threshold point in $\eta$ where adjacent Floquet bands touch (see Fig.~\ref{linear_lieb_bands_pi}) {and reopen with new gapless edge modes near the Floquet edges $\alpha_{edge} = \pm \pi/T$}. {Consequently, the Chern numbers are all zero and resemble those observed in \cite{Rudner2013} and \cite{Noh2017}.} The quasi one-dimensional rotation pattern {equation (\ref{1d_motion})--see Fig.(\ref{kagome_multi_band})(f)}
  is found to reverse the direction modes {propagate} {as well as the sign of the Chern invariants.}
  
{To} further highlight {these {similarities} among similar rotation patterns}, in Appendix \ref{honey_square_bands} we have included the Floquet bands for the honeycomb (see Fig.~\ref{honey_band_appendix}) and staggered square (see Fig.~\ref{square_band_appendix}) lattices for rotation patterns and parameters similar those considered above. Their bands are found to exhibit similar structure in response to {similar} driving patterns. {This highlights that lattice driving{,or combinations thereof,} could be tailored to suit the need of {potential} future applications involving these topologically protected modes.} 

\subsection{Kagome edge mode dynamics}

In this section we {provide} evolution dynamics for {several} 
 {Kagome edge} modes found in the previous section.
We integrate Eqs.~(\ref{kagome_TBA_eq1})-(\ref{kagome_TBA_eq3}) using the initial conditions
\begin{align}
\nonumber
a_{mn}(0)  & =  {\rm sech}\left(  \mu m \right) a_{n}(k_y) e^{i  m k_y } \; , \\ \label{linear_Kag_IC}
b_{mn}(0) & = {\rm sech}\left(  \mu m \right) b_{n}(k_y) e^{i  m k_y } \; , \\ \nonumber
c_{mn}(0) & = {\rm sech}\left(  \mu m \right) c_{n}(k_y) e^{i  m k_y }  \; ,
\end{align}
for the edge modes $a_n, b_n, c_n$ found by numerically solving system (\ref{1d_kagome_TBA_eq1})-(\ref{1d_kagome_TBA_eq3}) at a particular $k_y $. As with the Lieb lattice, we fix $\sum_n \left( |a_{n}|^2 + |b_{n}|^2 + |c_{n}|^2 \right)= 1.$ A slowly-varying ${\rm sech}$ envelope is attached in the $m$-direction with {$\mu = 0.1.$} Here we only show the most dominant sublattice edge mode profile either on the left (pointy edge) at $n=0$, or the right (straight edge) at $n = N \gg 1$.

Some 
 {evolution} patterns are shown in Fig.~\ref{kagome_evolve}.
The first case considered is that of same rotation among all sublattices (\ref{same_rot_func}) with corresponding Floquet bands shown in Fig.~\ref{kagome_same_rot_diff_BC}(b). 
 {Each gap has topologically protected edge modes on both sides, and each 
propagates with constant velocity.} 
By attaching a slowly-varying envelope along the edge initially (\ref{linear_Kag_IC}) we {typically excite just the desired mode {$k_y$ and a small sideband}. When the {spatial} envelope is too narrow we {find that} {many additional frequencies}}{, such as those that support bulk modes,} {can be excited.}

The evolutions in Figs.~\ref{kagome_evolve}(a) and (c) are both located along the pointy edge {and evolve at a constant negative velocity. The majority of the energy resides in the outermost column of lattice sites, located at {$b_{m,0} $}, hence we display the $b$-mode profiles.} 
 {Conversely,} along the straight edge [see Figs.~\ref{kagome_evolve}(b) and (d)] {the gapless modes propagate in the positive direction { and most of the energy is shared between the $a$ and $c$ sites{; we only} show the $c$-sublattice mode dynamics.}} 

The evolution of a Kagome lattice with a $\pi$-phase offset between the $b$ sites and the $a,c$ lattice sites (\ref{pi_phase_rot_func}) is shown in Figs.~\ref{kagome_evolve}(e-f). The corresponding dispersion curves are displayed in Fig.~\ref{kagome_multi_band}(d). The evolution dynamics here resemble those {of} the Lieb lattice [see Fig.~\ref{lieb_linear_evolve}(e-f)], as well as those seen in honeycomb and staggered square \cite{AbCo2017} when the sublattices are out of phase with each other. {In both cases shown here, the} modes are found to oscillate back-and-forth between the $a/c$ and $b$ lattice sites over the course of one lattice period ($T = 1.5$) and propagate with {the group velocity.} 

\begin{figure} [ht]
\centering
\includegraphics[scale=.6]{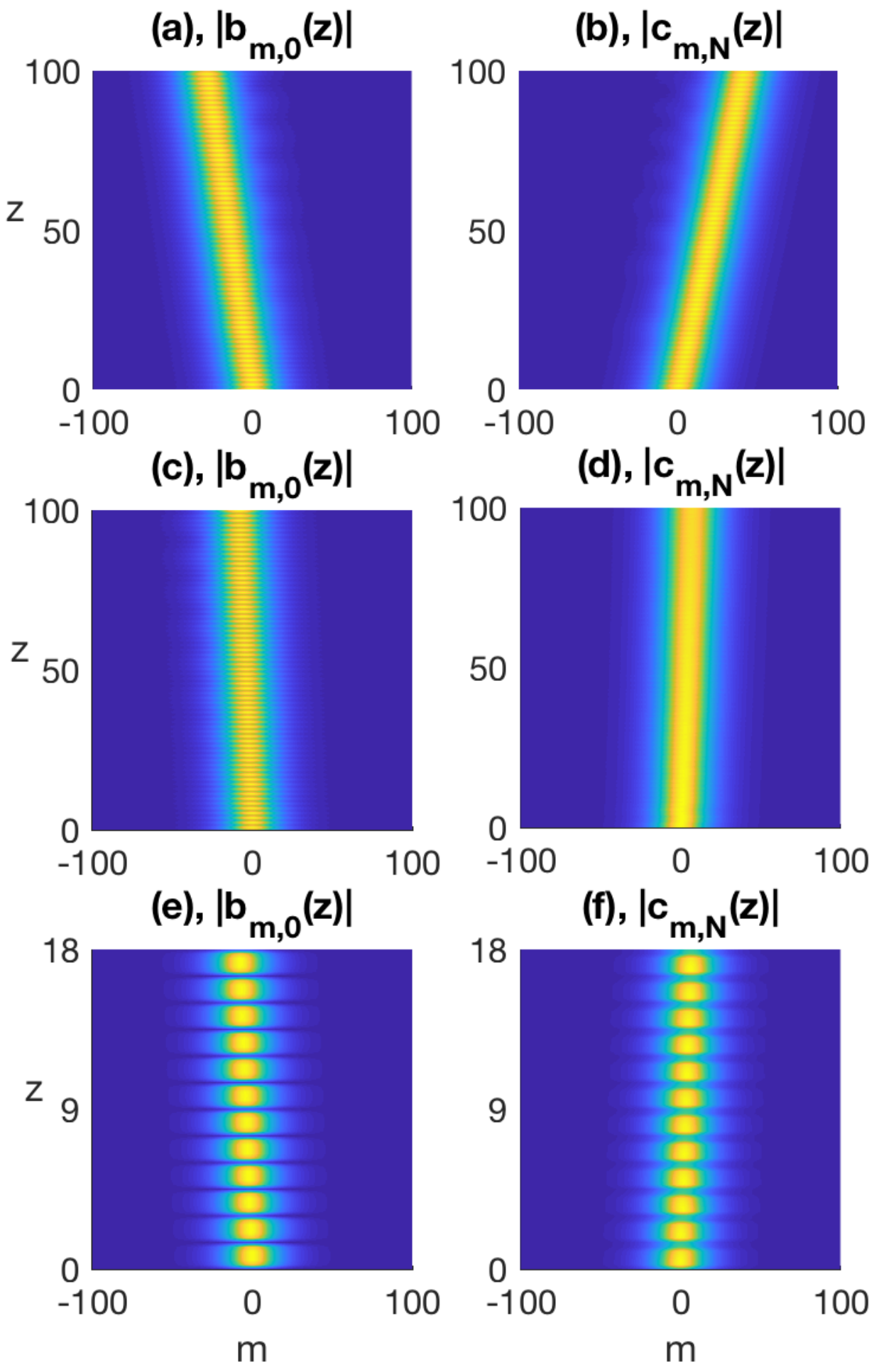}
\caption{Evolution of edge profile in the Kagome lattice. The edge modes shown in panels (a-d) correspond to the Floquet bands in Fig.~\ref{kagome_same_rot_diff_BC}(b) with Floquet exponents: {(a) $\alpha(1.2) = -.231$, (b) $ \alpha(1.2) =-.178$, (c) $ \alpha(1.2) =.343$, (d) $\alpha(3) =.453$.} The edge states in panels (e-f) correspond to the bands shown in  Fig.~\ref{kagome_multi_band}(d) at $k_y = 0.2$ {where $\alpha =$ (e) $1.84$ and (f) $2.04$.}} 
\label{kagome_evolve}
\end{figure}

\section{Defect Barrier}
\label{defect}

Edge states associated with other {cf.}\cite{Wang2009, RechtsSegev2013, AbCo2017} helically driven photonic lattices have been shown to exhibit robust {scatter-free} 
motion in the presence of lattice defects. 
Here we introduce a lattice defect barrier and monitor the mode evolution as it encounters this barrier. Physically, the defects we consider correspond to the absence of lattice site waveguides. 
As such, there is little beam propagation in these areas and so we {set} 
 {the} beam field to {be} zero there. 

\subsection{Linear Evolution}

First, consider the {linear} Lieb lattice with same rotation waveguide motion; the Floquet bands of which are shown in Fig.~\ref{linear_lieb_bands_same}(b). We evolve an edge state 
into a defect barrier {located in the region {$(m,n) \in [-30,-27]  \times  [0,2]$ along the bearded edge, and $(m,n) \in [27,30] \times [N-2,N]$ for $a,b$ sites and $[26,30] \times [N-2,N]$ for $c$ sites} along the straight edge. Put another way, the wave field is set to zero for every lattice site inside these regions}. In the absence of a barrier, the mode {profile} evolution {along the straight edge} {is} shown in {Fig.~\ref{lieb_linear_evolve}(b)}. Several snapshots {of {a} topologically protected mode} are displayed in the {right column} of Fig.~\ref{lieb_defect_evolve}. {All intensity plots given here (and below) are taken relative to the {first snapshot in the series.} 
{The topologically protected mode encounters the lattice defect and {we see that} it propagates around the barrier{; there is no backscatter}. We note that in several of our simulations if the mode envelope (\ref{linear_Lieb_IC}) is  too narrow i.e. $\mu$ not {sufficiently} small, then {modes outside the band gap {can be} excited and} noticeable deterioration of the edge state {occurs}. 



An example of a {Lieb} mode evolving through a defect along the bearded edge is shown in the {left column}  of Fig.~\ref{lieb_defect_evolve}. 
 {The rotation pattern used is that of $\pi$-offset among the sublattices (\ref{pi_phase_rot_func}) whose corresponding band diagram is given in Fig.~\ref{linear_lieb_bands_pi}(c). {In the absence of defect, the mode propagation is shown in Fig.~\ref{lieb_linear_evolve}(e).} The reason we choose to show a mode with this rotation pattern, and not same rotation, is because the group velocity is much larger [cf. Fig.~\ref{linear_lieb_bands_same}(b)], and we do not have to wait as long for the mode to evolve.}
Here the edge state is found to {also} pass scatter-free 
 through the defect.






\begin{figure} [ht]
\centering
\includegraphics[scale=.8]{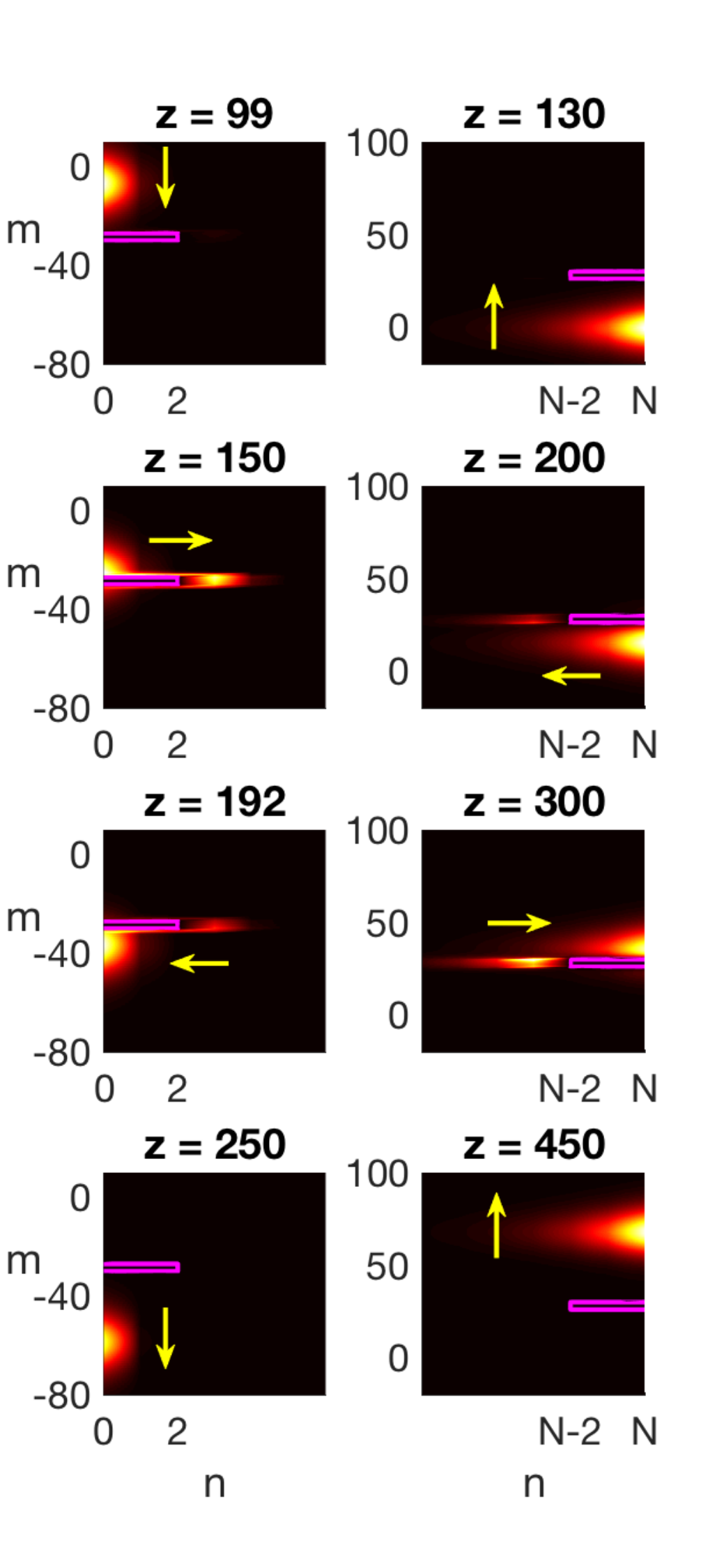}
\caption{{Linear} {topologically protected} edge mode dynamics encountering a defect barrier in the Lieb lattice. {(Left column)} Intensity evolution $|a_{mn}(z)|^2$ along bearded edge with $\pi$-offset rotation corresponding to the Floquet bands in Fig.~\ref{linear_lieb_bands_pi}(c) and {$(k_y,\alpha) = (0.2,-1.94)$}. {(Right column)} Intensity evolution $|c_{mn}(z)|^2$ along straight edge for same rotation corresponding to the Floquet bands in Fig.~\ref{linear_lieb_bands_same}(b) with $(k_y,\alpha) = (1.65,.0362)$. }
\label{lieb_defect_evolve}
\end{figure}

{We 
point out an interesting mode-defect interaction in the case of same sublattice rotation {(\ref{same_rot_func})} when the Floquet exponent {[see Fig.~\ref{linear_lieb_bands_same}(b)]} is not well separated from the bulk bands. 
The evolution of a mode along the straight edge corresponding to a Floquet exponent deep within the band gap {was} displayed in Fig.~\ref{lieb_defect_evolve} (right column). When we instead consider a mode, on the same Floquet edge band, 
 {whose position in the band digram {is nearby a bulk band}}
we observe significant {deterioration of the edge state}. In Fig.~\ref{lieb_barrier_bulk_scatter}, {for} a mode whose value $\alpha(k_y)$ lies nearby the bulk modes, {a significant portion} is observed to 
 {disperse} 
 away from the defect barrier and diminish in intensity. {We point out that this occurs despite the topological protection.} This mode appears to experience some transfer/leakage of energy into extended bulk states. {Topologically protected modes whose Floquet exponents {{are} located} deep in the gap {(well away from the bulk) appear to avoid this {effect.} }} 

\begin{figure} [ht]
\centering
\includegraphics[scale=.52]{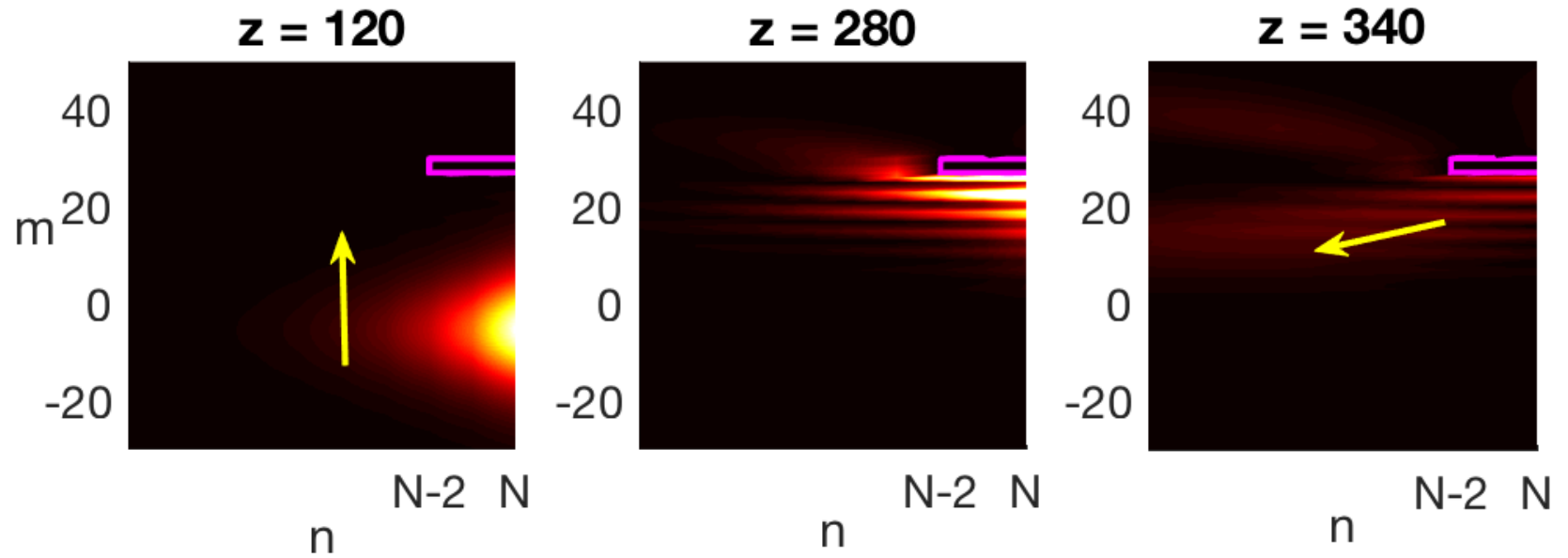}
\caption{{Linear} edge mode dynamics encountering a defect barrier in the Lieb lattice {with the Floquet band positioned close to the bulk.} Intensity evolution $|c_{mn}(z)|^2$ along straight edge for same rotation corresponding to the Floquet bands in Fig.~\ref{linear_lieb_bands_same}(b) at $\alpha(2) = .191$.}
\label{lieb_barrier_bulk_scatter}
\end{figure}

The mode evolutions for the Kagome lattice in the presence of a defect barrier are considered next.
 {The defect along the pointy edge lies in the region $0 \le n \le 3$, and $ N-3 \le n \le N$ along the straight edge, with $-1 \le m \le 2$ for $b$ and $c$ lattice sites and $-2 \le m \le 1$ for $a$ sites.}
Two intriguing cases are presented. The first is that of an edge mode propagating along a straight edge, {the} 
 {corresponding unidirectional} band diagram {is} shown in Fig.~\ref{kagome_same_rot_diff_BC}(b). In the absence of a barrier the mode propagation is displayed in Fig.~\ref{kagome_evolve}(b). The mode evolution with defect is {presented} in the {right column} of Fig.~\ref{kagome_defect_evolve}. The {topologically protected} edge state works its way around the defect and {exits with the intensity it entered with.} 


The next case we examine {has an edge mode with corresponding dispersion curve that is \it{not}} unidirectional throughout the Brillouin zone{, yet still crosses a band gap}; i.e. {a topological mode with group velocity that changes sign.} 
Consider the pointy edge mode located at $(k_y , \alpha) = (1.2,0.343)$ in  Fig.~\ref{kagome_same_rot_diff_BC}(b){; to the left of the minimum}. This band  has two different slope velocities signs depending on $k_y $. In the absence of defect, the {constant velocity} mode propagation is presented in Fig.~\ref{kagome_evolve}(c). 
With defect, the mode evolution is shown in the {left column} of Fig.~\ref{kagome_defect_evolve}. Here the mode travels towards the barrier, but instead of passing around the defect barrier, it ceases {movement in the negative direction and scatters backward with a noticeable loss {in} intensity}. This backscattering tends to occur when there are other modes, bulk or edge, at the same frequency $\alpha$ that can become excited at the defect. This was also the case for the mode shown in Fig.~\ref{lieb_barrier_bulk_scatter}.







\begin{figure} [ht]
\centering
\includegraphics[scale=.8]{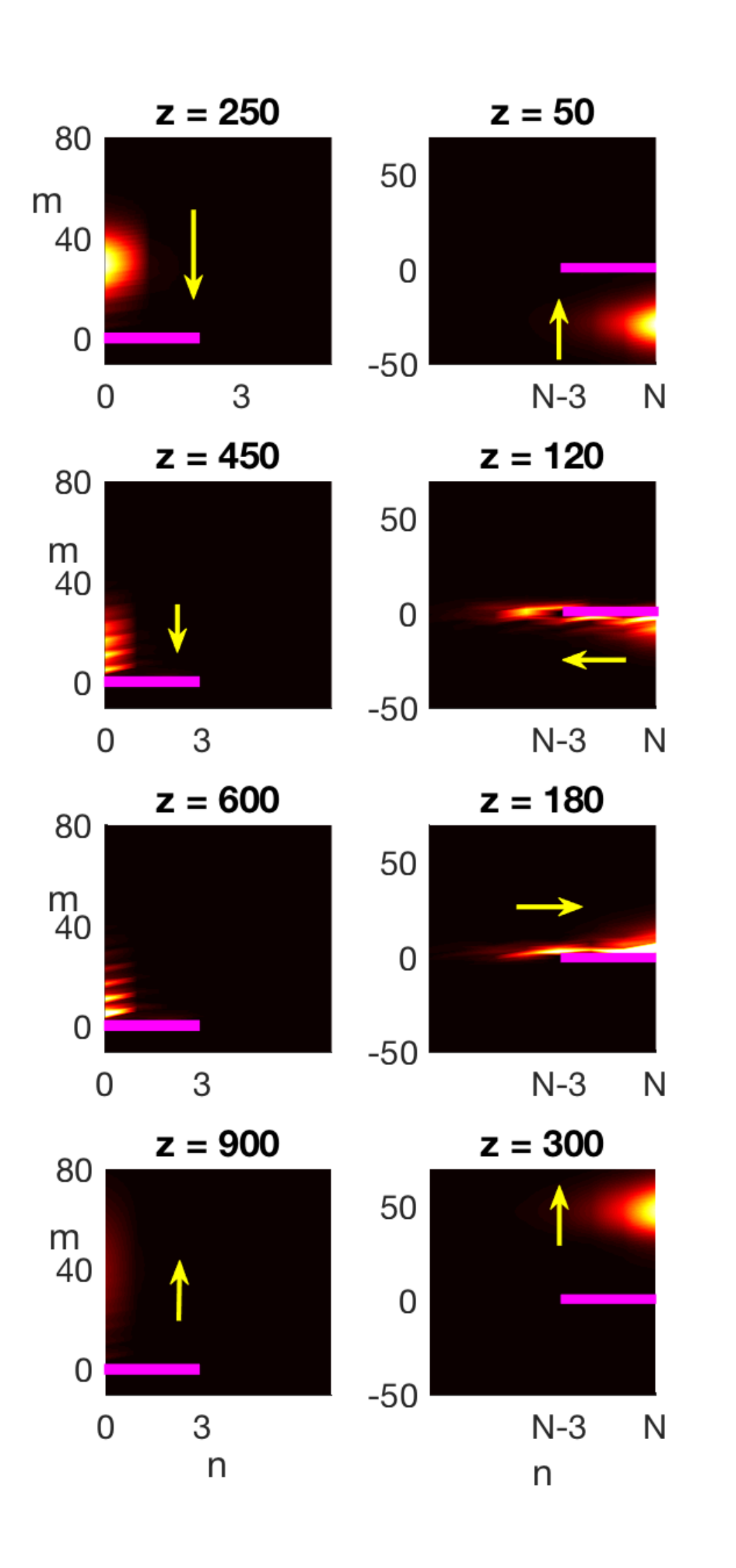}
\caption{{Linear} edge mode dynamics encountering a defect barrier in the Kagome lattice for {in-phase} rotation corresponding to the Floquet bands in Fig.~\ref{kagome_same_rot_diff_BC}(b). 
 {(Left column)} Intensity evolution $|b_{mn}(z)|^2$ along pointy edge with same rotation and {$(k_y,\alpha) = (1.2,.343)$}. {(Right column)} Intensity evolution $|c_{mn}(z)|^2$ along straight edge and {$(k_y,\alpha) = (1.2,-.178)$}.}
\label{kagome_defect_evolve}
\end{figure}


\subsection{Nonlinear Evolution}

{In this {final} section we consider nonlinear modes ($\gamma_{\rm nl} \not= 0 $) in the presence of lattice defects. For weak nonlinearity, with parameters similar to those considered above, nonlinear modes {in honeycomb lattice have been} found to exist {when there is longitudinal driving  \cite{AbCuMa2014,AbCo2017}.} 
There it was found that the edge solitons {`inherited'} the topological protection from the linear mode; we look to see if that is also the case here.}


{To begin, we {note that there is} a small parameter  that is related to the {relatively rapid} driving frequency $\epsilon = 1/ \Omega \approx 0.239$. The weak nonlinearity {coefficient} is set to $\gamma_{\rm nl} = \epsilon.$ We next look for a soliton mode, which is a balance between dispersion and self-focusing nonlinearity. As such, we initialize these modes similar to (\ref{linear_Lieb_IC}) and (\ref{linear_Kag_IC}) except we add an amplitude {coefficient}  $A$ in front of the {hyperbolic secant} terms  that satisfy the balance $A^2 \gamma_{\rm nl} = \mu^2 |\alpha''(k_y)| $. 
 {The full {\it nonlinear} Lieb (\ref{lieb_TBA_eq1})-(\ref{lieb_TBA_eq3}) and Kagome (\ref{kagome_TBA_eq1})-(\ref{kagome_TBA_eq3}) systems are {then} evolved in $z$.}


\begin{figure} [ht]
\centering
\includegraphics[scale=.8]{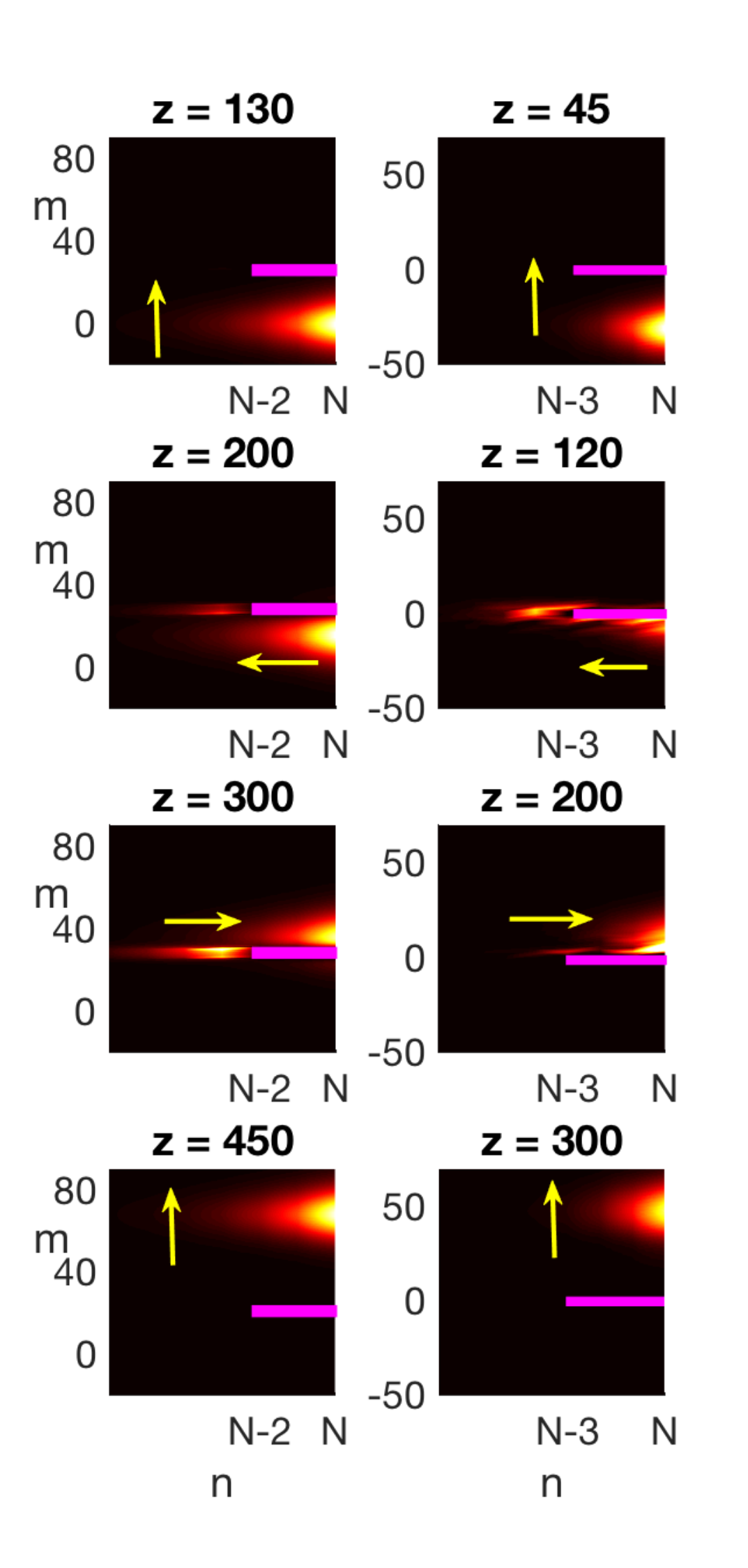}
\caption{{Nonlinear ($\gamma_{\rm nl} = \epsilon = 1/\Omega$) edge mode dynamics encountering a defect barrier. (Left column) Intensity, $|c_{mn}(z)|^2$, in Lieb lattice with same parameters as those given in right column of Fig.~\ref{lieb_defect_evolve} {with} dispersion $  \alpha''(1.65) = -.0364$ {corresponding to Fig.~\ref{linear_lieb_bands_same}(b).} (Right column) Intensity, $|c_{mn}(z)|^2$, in Kagome lattice with same parameters as those given in right column of Fig.~\ref{kagome_defect_evolve} {with} dispersion $\alpha''(1.2) = .141$ {corresponding to Fig.~\ref{kagome_same_rot_diff_BC}(b).}}}
\label{nonlinear_defect_evolve}
\end{figure}

In Fig.~\ref{nonlinear_defect_evolve} we show two {nonlinear} topologically protected {solitons:} 
one {with in Lieb lattice and  and the other Kagome,} using the same parameters taken in Figs.~\ref{lieb_defect_evolve} and \ref{kagome_defect_evolve} {along the straight edge}. 
{Taking the balance given above,} the nonlinear modes are found to closely mirror the unidirectional behavior of the linear states. Namely, for edge modes whose {Floquet exponent is} deep in the gap and {{slope-definite}} {throughout the gap} we observe scatter-free propagation {around the defect}. When the mode is not {sufficiently} deep in the gap or the sign of the group velocity depends on {$k_y$}, then the nonlinear mode can {exhibit backscatter} (see Figs.~\ref{lieb_barrier_bulk_scatter} and \ref{kagome_defect_evolve}). These nonlinear states are appealing since they 
combine topologically protected one-way motion with the robustness and balance of solitons.}

\section{Conclusions}
\label{conclude}

Tight-binding approximations {that} describe deep longitudinally driven Lieb and Kagome waveguide lattices {were constructed}. These lattices were decomposed into three sublattices such that each sublattice moves with its own driving pattern e.g. phase offset, different radii, different frequency, etc. 
We considered periodically oscillating waveguides and computed their corresponding Floquet bands. These dispersion bands were found to support localized {topologically protected} edge modes that were either stationary (flat band) or traveling. {A topological invariant, the  Chern number was calculated; this demonstrates that there is nontrivial associated topology present in the {dispersion bands}.} 

{Topologically protected} modes {located in the band gap} with sign-definite group velocity {throughout the gap} were found to {propagate unidirectionally}  around lattice defect barriers, and did not significantly backscatter. 
If, on the other hand, the corresponding Floquet band has sign-indefinite group velocity, then the edge mode could {backscatter at} the barrier. {Additionally, when the corresponding Floquet exponent {is} not well-separated from the bulk band{, then} {strong dispersion} {could occur}.} {These observations highlight {further insights regarding}  topologically protected modes.}

{Two {additional} conclusions can be drawn from this work: (i) Numerous {physically significant} lattices can be used to find topologically protected edge {modes} e.g. honeycomb, {staggered} square, Lieb, {Kagome.} The tight-binding approach we have developed 
 {can} be generalized to incorporate  {more} detailed {features} e.g. next-nearest neighbor interaction, detuning in the refractive index, etc. (ii) Many different types of longitudinal driving can generate {topologically protected} 
 edge modes. {Moreover,} different rotation patterns can be selected to suit {applications} 
 e.g. along the bearded edge of a Lieb lattice $\pi$-offset rotation creates faster propagating modes than in-phase rotation.}

{Finally, weakly nonlinear modes were found to inherit the topological protection of the linear Floquet modes. These modes present an exciting opportunity to merge the one-way properties of the linear problem with {intense} nonlinear states i.e. solitons.}

\section{Acknowledgements}
This work was partially supported by AFOSR under grant No. FA9550-16-1-0041.

\bibliography{Lieb_Kagome_Mnscrpt_3_8_19.bib}

\appendix

\section{Tight-binding Model Coefficients}
\label{TBA_coefficients}

{The} coefficients in the tight-binding models given in Sec.~\ref{TBA_derive} are presented here. The Lieb (\ref{lieb_TBA_eq1})-(\ref{lieb_TBA_eq3}) and Kagome (\ref{kagome_TBA_eq1})-(\ref{kagome_TBA_eq3}) systems are both defined in terms of the {distance-dependent coefficient $\mathbb{L}({\bf v})$}. The only difference in the two tight-binding models is the individual lattice configurations i.e. the different lattice vectors between nearest neighbor lattice sites, and {the driving pattern being used.}
 {We point out that we neglect {a small imaginary} coefficient $\mathbb{R}_j$ that was used in \cite{AbCo2017}. The reason is that typically $V_0$ {is} large and so $|\mathbb{L}({\bf v})| \gg 1$ is the {asymptotically} dominant term. {As a result, the numerically computed spectrum of the Lieb and Kagome systems in Sec.~\ref{TBA_derive} is found to be real.}} 

The linear coefficient in the tight-binding approximations is
\begin{align}
\label{linear_coeff_define}
&\mathbb{L}({\bf v} - \Delta {\bf h}_{ij}(z)) =  \bigg[ V_0^3  \sqrt{\frac{\sigma_x \sigma_y}{(1 + \sigma_x V_0) (1 + \sigma_y V_0 )}} \\ \nonumber
&  \times \bigg( 2 e^{- \frac{V_0}{4} \left[ \frac{[{\bf v} - \Delta {\bf h}_{ij}(z)]_x^2}{\sigma_x(1 + V_0 \sigma_x)} + \frac{[{\bf v} - \Delta {\bf h}_{ij}(z)]_y^2}{\sigma_y (1 + V_0 \sigma_y)} \right]} - 1 \bigg)   \\ \nonumber
&+  \frac{V_0^2}{4} \bigg( \frac{[{\bf v} - \Delta {\bf h}_{ij}(z)]_x^2}{\sigma_x^2} + \frac{[{\bf v} - \Delta {\bf h}_{ij}(z)]_y^2}{\sigma_y^2} \bigg) \bigg] \\ \nonumber
& \times e^{- \frac{V_0}{4} \left( \frac{[{\bf v} - \Delta {\bf h}_{ij}(z)]_x^2}{\sigma_x} + \frac{[{\bf v} - \Delta {\bf h}_{ij}(z)]_y^2}{\sigma_y} \right)} e^{i {\bf v}\cdot {\bf A}(z) } \; ,
\end{align}
where the $x$ and $y$ subscripts denote the {first and second} components, respectively.
The relative driving motion is captured by the functions  $\Delta {\bf h}_{ij}(z) \equiv {\bf h}_i(z) - {\bf h}_j(z)$ for $i,j = a,b,c$.
The defining lattice vectors for the Lieb and Kagome lattices are given in Eq.~(\ref{lieb_lattice_vec_define}) and (\ref{kagome_lattice_vec_define}), respectively.
The nonlinearity coefficient for the tight-binding models is given by
\begin{equation*}
\gamma_{\rm nl} = \frac{\gamma  V_0}{2 \pi \sqrt{\sigma_x \sigma_y}} \ge 0 \; .
\end{equation*}

\section{Computation of Chern Numbers}
\label{chern_section}

{Bulk bands with associated nontrivial Chern numbers are known to support gapless edge modes \cite{Lu2014}.  The Chern number for a bulk band is equal to the number of gapless modes above the band minus the number of gapless modes below it.  {The orientation of the chirality must also be taken into account: modes moving in the counter-clockwise (clockwise) direction have a positive (negative) chirality. This is reflected in the Chern numbers, in particular in the quasi 1d motion case (\ref{1d_motion}), where the edge modes travel  clockwise around the boundary and the corresponding Chern numbers have opposite sign.} {Moreover, the Chern invariants in the spectral problem [see Eq.~(\ref{spectral_sys})] are the same as those in the edge problem.}  {When a defect along the boundary is introduced the Chern invariants are unaffected and a topologically protected mode will persist.} 
This is the {Bulk-Edge} correspondence. Hence our interests are served by computing the Chern invariants in order to establish whether or not the modes we find are indeed topologically protected.

The Chern number of the bulk (spectral) bands is computed directly from the Lieb (\ref{lieb_TBA_eq1})-(\ref{lieb_TBA_eq3}) and Kagome (\ref{kagome_TBA_eq1})-(\ref{kagome_TBA_eq3}) discrete systems. {For both systems we take plane wave solutions of the form 
\begin{align}
\label{plane_wave_solns}
& a_{mn} = \delta({\bf k},z) e^{i {\bf k} \cdot (m {\bf w}_1 + n {\bf w}_2)} \; , \\ \nonumber
& b_{mn} = \beta({\bf k},z) e^{i {\bf k} \cdot (m {\bf w}_1 + n {\bf w}_2)} \; , \\ \nonumber
& c_{mn} = \gamma({\bf k},z) e^{i {\bf k} \cdot (m {\bf w}_1 + n {\bf w}_2)} \; ,
\end{align}
where {${\bf w}_1 = 2 {\bf e}_2, {\bf w}_2 = 2{\bf e}_1$} for the Lieb lattice and ${\bf w}_1 = {\bf e}_2, {\bf w}_2 = \sqrt{3}{\bf e}_1$ for Kagome. Substituting (\ref{plane_wave_solns}) into the tight-binding equations yields a $3\times 3$ coupled system of equations
\begin{equation}
\label{spectral_sys}
 \frac{d {\bf c}}{d z} = i \mathcal{M}({\bf k} ,z) {\bf c} \; , ~~~~~ {\bf c} = (\delta , \beta ,\gamma)^T \; ,
\end{equation}
such that {
\begin{widetext}
$$ \mathcal{M}_{\rm Lieb}({\bf k} ,z) = 
 \begin{pmatrix}
\Delta {\bf h}_{ab} \cdot {\bf A}_z & \mathbb{L}_1^{ab}(z) + \mathbb{L}_{-1}^{ab}(z) e^{-i {\bf k} \cdot 2 {\bf e}_1} & 0 \\
\mathbb{L}_1^{ba}(z) e^{i {\bf k} \cdot 2 {\bf e}_1} + \mathbb{L}_{-1}^{ba}(z)  & 0 &  \mathbb{L}_2^{bc}(z)  + \mathbb{L}_{-2}^{bc}(z) e^{-i {\bf k} \cdot 2 {\bf e}_2}  \\
0 & \mathbb{L}_2^{cb}(z) e^{i {\bf k} \cdot 2 {\bf e}_2} + \mathbb{L}_{-2}^{cb}(z)  & \Delta {\bf h}_{cb} \cdot {\bf A}_z
\end{pmatrix} \; , $$ 
\end{widetext}
}for the Lieb lattice, where $\mathbb{L}_{\pm l}^{ij}(z) = \mathbb{L}(\pm {\bf e}_l - \Delta {\bf h}_{ij}(z))$, and in the case of the Kagome lattice, the matrix is
\begin{widetext}
$$ \mathcal{M}_{\rm Kagome}({\bf k} ,z) = 
 \begin{pmatrix}
\Delta {\bf h}_{ab} \cdot {\bf A}_z & \mathbb{L}_{-1}^{ab}(z)  + \mathbb{L}_{1}^{ab}(z) e^{i {\bf k} \cdot ({\bf w}_1 + {\bf w}_2)} & \mathbb{L}_{-3}^{ac}(z)  + \mathbb{L}_{3}^{ac}(z) e^{i {\bf k} \cdot (2 {\bf w}_1 )} \\
 \mathbb{L}_{-1}^{ba}(z) e^{-i {\bf k} \cdot ({\bf w}_1 + {\bf w}_2)} + \mathbb{L}_{1}^{ba}(z)  & 0 & \mathbb{L}_{-2}^{bc}(z) e^{-i {\bf k} \cdot ( {\bf w}_2 - {\bf w}_1 )} + \mathbb{L}_{2}^{bc}(z)   \\
\mathbb{L}_{-3}^{ca}(z) e^{-i {\bf k} \cdot (2 {\bf w}_1 )}  + \mathbb{L}_{3}^{ca}(z) & \mathbb{L}_{-2}^{cb}(z) + \mathbb{L}_{2}^{cb}(z) e^{i {\bf k} \cdot ( {\bf w}_2 - {\bf w}_1 )}  & \Delta {\bf h}_{cb} \cdot {\bf A}_z
\end{pmatrix} \; ,$$ 
\end{widetext}
where here $\mathbb{L}_{\pm l}^{ij}(z) = \mathbb{L}(\pm {\bf v}_l - \Delta {\bf h}_{ij}(z))$ for the vectors ${\bf v}_l, l = 1,2,3$ given in Eq.~(\ref{kagome_lattice_vec_define}).}
The eigenmode solutions are computed via Floquet theory in a manner similar to that described {above} Eq.~(\ref{floquet_exponent}). The 2d Chern number for the $p^{\rm th}$ spectral band is defined by the following integral over the Brillouin zone (BZ)
\begin{equation}
\label{chern_define}
C_p = \frac{1}{2\pi i} \iint_{BZ} (\nabla_{\bf k} \times {\bf A}_p) \cdot {\widehat k} ~d{\bf k} , 
\end{equation}
in terms of the Berry connection 
 {$ {\bf A}_p({\bf k},z) =  {\bf c}_p^\dag(z; {\bf k}) \partial_{k_x}  {\bf c}_p(z; {\bf k})  \ihat + {\bf c}_p^\dag (z; {\bf k}) \partial_{k_y}  {\bf c}_p(z; {\bf k})  \jhat $,}
 {for the eigenmode ${\bf c}_p(z; {\bf k})$ corresponding to the $p^{\rm th} $ spectral band,} evaluated at any $z$ {(see Appendix \ref{chern_conserve})}, where $\nabla_{\bf k}$ is the gradient in ${\bf k}$, and $^\dag$ denotes the complex conjugate transpose. Here we evaluate the Chern number at $z = 0.$ 


To numerically calculate the Chern numbers (\ref{chern_define}) the algorithm proposed in \cite{Fukui2005} is used {and summarized below}. Consider a square lattice with period $\ell$. Any non-square lattice can be mapped to a square lattice by a linear transformation. In the algorithm, a square Brillouin zone $[0,2\pi/\ell) \times [0,2\pi/\ell) $ is discretized by 
\begin{equation}
{\bf k}_{rs} = \left( \frac{2 \pi r }{\ell N} , \frac{2 \pi s}{\ell N} \right) \; , ~~~~ r,s = 0,1,\dots N-1 \; .
\end{equation}
The so-called unitary link variable is given by
\begin{equation}
U_{\mu,\nu}({\bf k}_{rs}) =  \frac{{\bf c}_p^*({\bf k}_{rs}) {\bf c}_p({\bf k}_{r+\mu,s+\nu}) }{\left| {\bf c}_p^*({\bf k}_{rs}) {\bf c}_p({\bf k}_{r+\mu,s+\nu})  \right|} \; ,
\end{equation}
where $\mu , \nu = 0, 1$. Normalizing the link variables this way isolates the phase.
The Berry curvature at the point ${\bf k}_{rs}$ is approximated by
\begin{equation}
F({\bf k}_{rs}) = \ln\left[ \frac{U_{1,0}({\bf k}_{rs}) U_{0, 1}({\bf k}_{r+1,s}) }{U_{1,0}({\bf k}_{r,s+1}) U_{0,1}({\bf k}_{rs}) } \right] \; .
\end{equation}
Finally, the Chern number integral in (\ref{chern_define}) is approximated by the sum
\begin{equation}
\tilde{C}_p = \frac{1}{2 \pi i} \sum_{r,s}  F({\bf k}_{rs}) \; .
\end{equation}
 In the limit $ N \rightarrow \infty$ it can be shown that $\tilde{C}_p \rightarrow C_p.$
}

\section{Floquet Bands for Honeycomb and Staggered Square Lattices}
\label{honey_square_bands}

{In our studies we have observed that {quite often} similar sublattice driving patterns} yield similar Floquet band {structure}, regardless of the underlying lattice. 
To emphasize this point we {will show typical} dispersion bands {corresponding to} 
 the honeycomb and staggered square lattices \cite{AbCo2017} for rotation parameters similar to those considered above. These lattices possess two lattice sites per unit cell, unlike the three lattice sites per unit cell for Lieb and Kagome. In terms of the discrete system{s} given in Sec.~\ref{TBA_derive}, these simpler systems {are written in a similar form {but with a equations instead: see \cite{AbCo2017}}.}
 {Moreover, we compute the Chern numbers to establish {that there are} topologically protected edge modes.} {One difference between the bands computed in \cite{AbCo2017} and those shown here is that we neglect the $\mathbb{R}_j$ term in the linear coefficients i.e. {$\mathbb{L}_+({\bf v}) \rightarrow \mathbb{L}({\bf v})$}. The justifications are discussed in Appendix \ref{TBA_coefficients}.}
 
 
\begin{figure} [ht]
\centering
\includegraphics[scale=.64]{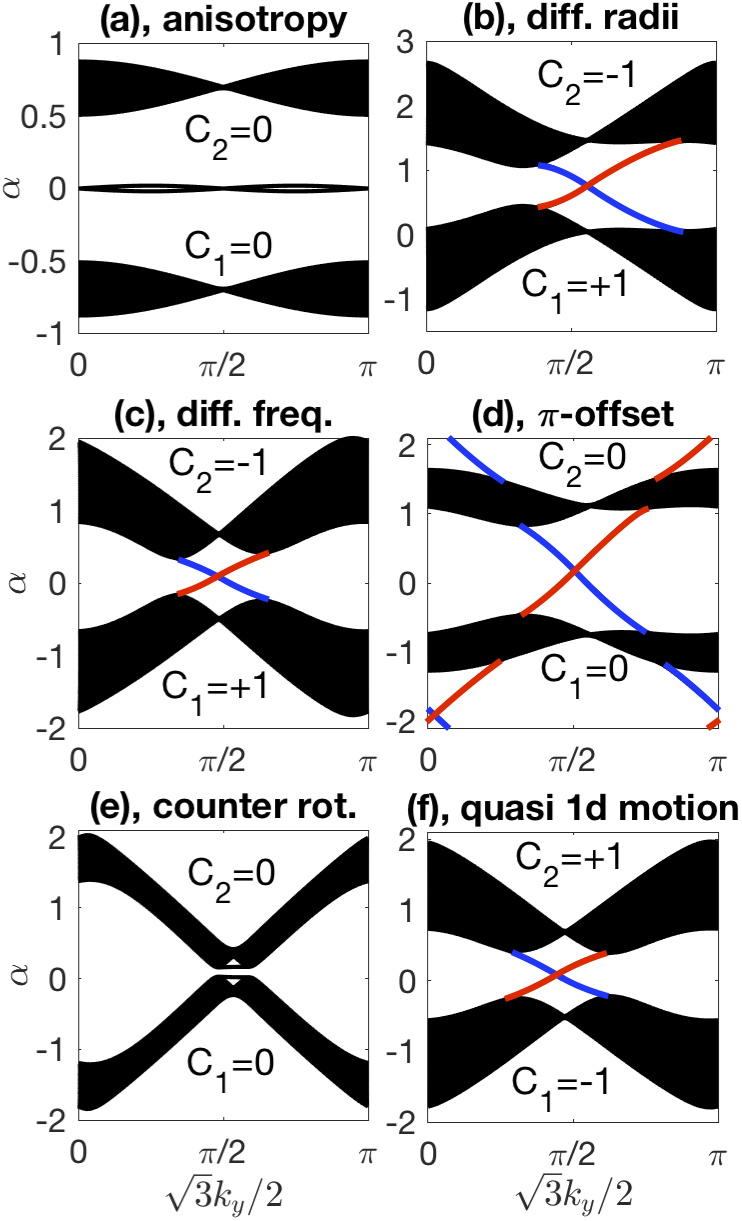}
\caption{(Color online) Honeycomb Floquet bands for various rotation patterns with zig-zag boundary conditions. Red curves indicate {gapless} edge modes on right edge, blue curves denote left {gapless} edge modes. The parameters used are the same as in Fig.~\ref{kagome_multi_band} with $a_{mn} (z)= 0$, {except (e) where $\eta = 2.2/15$}.} 
\label{honey_band_appendix}
\end{figure}

The Floquet bands for {a typical} honeycomb lattice are shown in Fig.~\ref{honey_band_appendix}. The parameters are identical to those used in Fig.~\ref{kagome_multi_band}{, except where noted}. In the case of anisotropy [Fig. \ref{honey_band_appendix}(a)], the non-topological edge bands do not span the gap and the Chern numbers are trivial. Topologically protected traveling modes are observed in the case of different radii [Fig.~\ref{honey_band_appendix}(b)], different frequency [Fig.~\ref{honey_band_appendix}(c)], $\pi$-offset [Fig.~\ref{honey_band_appendix}(d)] and quasi 1d motion [Fig.~\ref{honey_band_appendix}(f)]. {Since there are the same number of chiral modes above and below the $\pi-$offset bulk bands, the corresponding Chern numbers are zero.} The edge modes found in the case of counter rotation [see Fig.~\ref{honey_band_appendix}(e)] do not {span the gap or travel i.e. $\alpha'(k_y) = 0.$} 

Similar to the Lieb and Kagome lattices, the $\pi$-offset rotation has an additional family of solutions located at the Floquet edge.
 {Unlike the other lattices, the honeycomb lattice possesses non-topological modes in the case of anisotropy.} The modes for quasi 1d motion are found to travel in propagate in the opposite direction relative to all other rotation patterns, just like the Lieb, Kagome and square lattices (see below). {Since the edge modes travel with opposite chirality, the sign of the corresponding Chern numbers is also inverted: the lower (upper) bulk band has a Chern number of $-1 (+1).$}



The next set of band diagrams are for the staggered square lattice (see Fig.~\ref{square_band_appendix}).
In this case the underlying lattice is actually simple {when the two sublattices rotate in phase with each other}.  As a result, for in-phase rotation with anisotropic waveguides we obtain the bands in Fig.~\ref{square_band_appendix}(a) that {only have one bulk band and do} not support any edge modes. 

By driving two sublattices in different patterns 
 topological edge modes can be generated. 
 Traveling unidirectional{/topological} modes are generated by different radii [Fig.~\ref{square_band_appendix}(b)], different frequency [Fig.~\ref{square_band_appendix}(c)], $\pi$-offset [Fig.~\ref{square_band_appendix}(d)], counter rotation [Fig.~\ref{square_band_appendix}(e)], and quasi 1d motion [Fig.~\ref{square_band_appendix}(f)]. 
 {The different frequency [Fig.~\ref{square_band_appendix}(c)] and $\pi$-offset [Fig.~\ref{square_band_appendix}(d)] bands possess {a parameter regime with a }  ``mini-gap'' that {is} 
 numerically challenging to resolve. As a result, we do not include their Chern numbers, but do note that the modes in the upper and lower gaps exhibit robust unidirectional motion characteristic of topologically protected states.}

Like each other case considered above, the $\pi$-offset rotation pattern has a radius threshold necessary to generate the modes shown in Fig.~\ref{square_band_appendix}(d). 
For the different radii bands shown in Fig. 17(b) additional nontopological modes are found near $\alpha = 1$, {and the Chern numbers reflect this}. {The central gap modes in counter rotation, and quasi 1d motion do span the gap and are topologically protected, dissimilar to the honeycomb lattice above.} Similar to the Lieb, Kagome, and honeycomb lattices, modes generated from the quasi 1d motion bands propagate in an opposite orientation {i.e. chirality}, relative to most other cases.


\begin{figure} [ht]
\centering
\includegraphics[scale=.62]{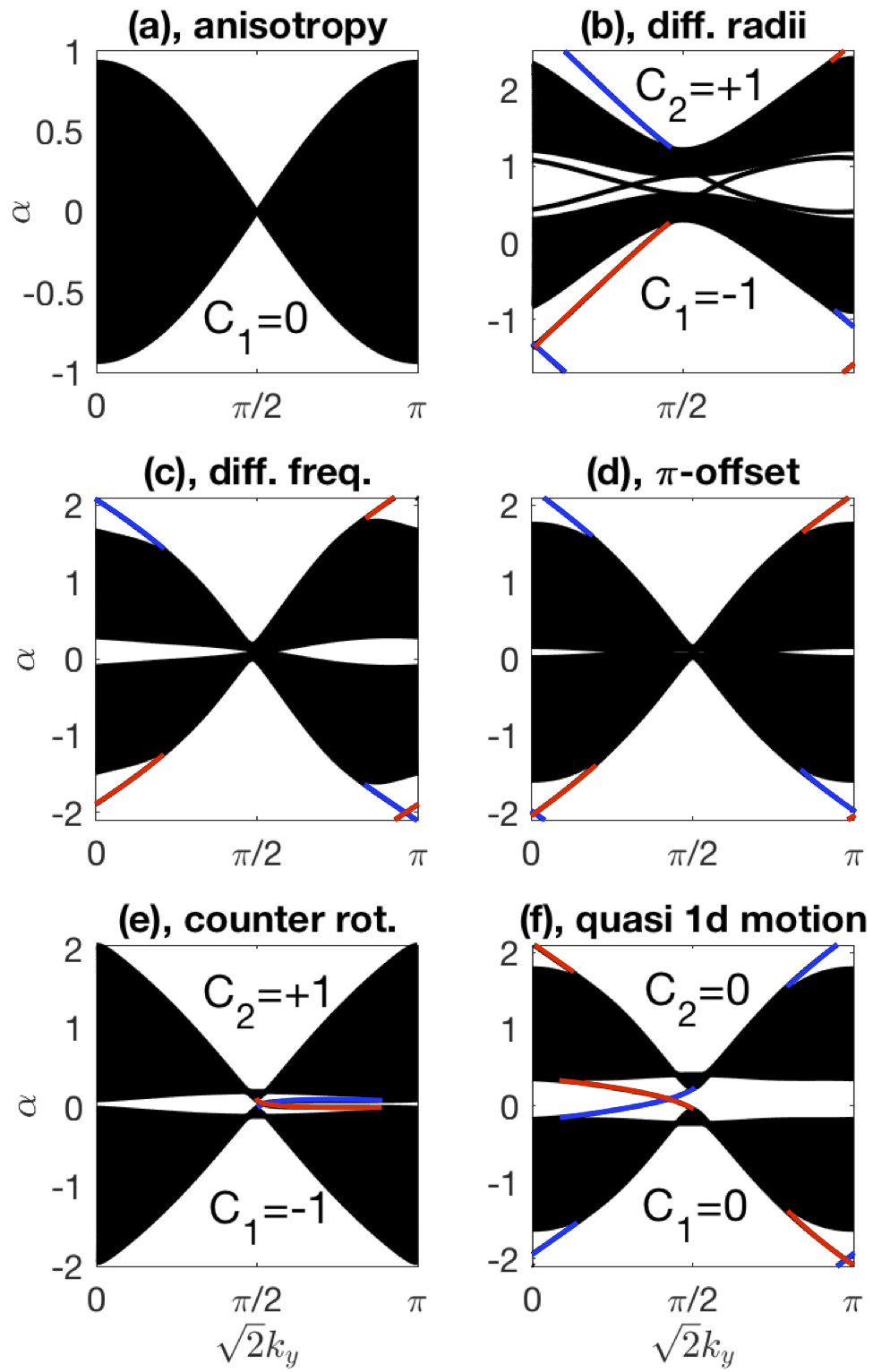}
\caption{(Color online) Staggered square Floquet bands for various rotation patterns. Red curves indicate gapless edge modes on right edge, blue curves denote left gapless edge modes. The parameters used are the same as in Fig.~\ref{kagome_multi_band} with $a_{mn} (z)= 0$, {except (d) and (e) where $\eta = 1.5/15.$}} 
\label{square_band_appendix}
\end{figure}

{
\section{Chern Number Invariance}
\label{chern_conserve}

In Appendix \ref{chern_section} the Chern number (\ref{chern_define}) and Bulk-Edge correspondence were computed and discussed. The eigenfunctions in (\ref{spectral_sys}) depend on $z$ (in general) and it is not clear that the Chern number is an invariant quantity i.e. $C(z) = {\rm constant}$. The purpose of this appendix is to prove that the Chern number is a $z$-independent quantity by showing $\frac{d C }{dz} = 0.$ We show this for the Lieb lattice (the Kagome lattice follows in a similar manner).

Recall, for system (\ref{spectral_sys}), the 2d Chern number is
\begin{equation}
\label{chern2}
C = \frac{1}{2 \pi i} \iint_{BZ} \left(  \frac{\partial {\bf c}^{\dag}}{\partial k_x}  \frac{\partial {\bf c}}{\partial k_y}- \frac{\partial {\bf c}^{\dag}}{\partial k_y}  \frac{\partial {\bf c}}{\partial k_x} \right)  dk_x dk_y \; ,
\end{equation}
where $\dag$ denotes the complex conjugate transpose.
One key observation is the symmetry in the linear coefficients (\ref{linear_coeff_define}):
\begin{equation}
\label{L_symmetry}
\mathbb{L}_{- \ell}^{ij} = \mathbb{L}(-{\bf v}_{\ell} - \Delta {\bf h}_{ij}) = \mathbb{L}^*({\bf v}_{\ell} - \Delta {\bf h}_{ji}) = \left( \mathbb{L}_{ \ell}^{ji}  \right)^* \; ,
\end{equation}
where $*$ denotes complex conjugation. The Brillouin zone for the Lieb lattice in Fig.~\ref{lieb_fig} is the square $[0, \pi] \times [0 , \pi].$ The Lieb coefficient matrix below (\ref{spectral_sys}) is periodic $\mathcal{M}(k_x + \pi ,k_y  ,z) = \mathcal{M}(k_x,k_y,z)  = \mathcal{M}(k_x  ,k_y + \pi ,z) $ and,  by symmetry (\ref{L_symmetry}), Hermitian: $\mathcal{M}^\dag = \mathcal{M}$.

 Differentiating (\ref{chern_define}) with respect to $z$ yields
\begin{align}
\label{dCdz_eqn}
\frac{d C}{dz} = \frac{1}{2 \pi} \iint_0^{\pi} &\bigg( \frac{\partial {\bf c}^\dag}{\partial k_x} \frac{\partial \mathcal{M}}{\partial k_y} {\bf c} + {\bf c}^\dag \frac{\partial \mathcal{M}}{\partial k_y} \frac{\partial {\bf c}}{\partial k_x} \\ \nonumber
- &{\bf c}^\dag \frac{\partial \mathcal{M}}{\partial k_x} \frac{\partial {\bf c}}{\partial k_y} - \frac{\partial {\bf c}^\dag}{\partial k_y} \frac{ \partial \mathcal{M}}{\partial k_x} {\bf c}  \bigg) dk_x dk_y \; .
\end{align}
{We assume} that if the eigenfunction ${\bf c}$ is periodic in ${\bf k}$ {at $z=0$: ${\bf c}(k_x + \pi, k_y, z=0) = {\bf c}(k_x, k_y, z=0) = {\bf c}(k_x, k_y + \pi, z=0),$ then ${\bf c}$ remains periodic in ${\bf k}$ for all $z$.} 
Applying integration-by-parts on (\ref{dCdz_eqn}), along with $\frac{\partial^2 \mathcal{M}}{\partial k_x \partial k_y} = 0,$ gives the desired result. Hence the Chern numbers in Eq.~(\ref{chern_define}) are independent of $z$.
}

\end{document}